\documentclass[fleqn,usenatbib]{mnras}

\usepackage{newtxtext,newtxmath}

\usepackage[T1]{fontenc}
\usepackage{graphicx}
\usepackage{subcaption}
\usepackage{siunitx}

\DeclareRobustCommand{\VAN}[3]{#2}
\let\VANthebibliography\thebibliography
\def\thebibliography{\DeclareRobustCommand{\VAN}[3]{##3}\VANthebibliography}

\usepackage{graphicx}	
\usepackage{amsmath}	

\usepackage{scalerel,tikz}
\usetikzlibrary{svg.path}
\definecolor{orcidlogocol}{HTML}{A6CE39}
\tikzset{orcidlogo/.pic={\fill[orcidlogocol] svg{M256,128c0,70.7-57.3,128-128,128C57.3,256,0,198.7,0,128C0,57.3,57.3,0,128,0C198.7,0,256,57.3,256,128z}; \fill[white] svg{M86.3,186.2H70.9V79.1h15.4v48.4V186.2z} svg{M108.9,79.1h41.6c39.6,0,57,28.3,57,53.6c0,27.5-21.5,53.6-56.8,53.6h-41.8V79.1z M124.3,172.4h24.5c34.9,0,42.9-26.5,42.9-39.7c0-21.5-13.7-39.7-43.7-39.7h-23.7V172.4z} svg{M88.7,56.8c0,5.5-4.5,10.1-10.1,10.1c-5.6,0-10.1-4.6-10.1-10.1c0-5.6,4.5-10.1,10.1-10.1C84.2,46.7,88.7,51.3,88.7,56.8z};}}
\newcommand\orcidicon[1]{\href{https://orcid.org/#1}{\mbox{\scalerel*{
\begin{tikzpicture}[yscale=-1,transform shape]\pic{orcidlogo};
\end{tikzpicture}}{|}}}}

\title[Collisionless turbulent dynamo]{Critical magnetic Reynolds number of the turbulent dynamo in collisionless plasmas}

\author[Achikanath Chirakkara et al.]{
Radhika Achikanath Chirakkara,$^{\orcidicon{0000-0001-5583-5038}\,1}$\thanks{E-mail: \href{mailto:radhika.achikanathchirakkara@anu.edu.au}{radhika.achikanathchirakkara@anu.edu.au}}
Amit Seta,$^{\orcidicon{0000-0001-9708-0286}\,1}$
Christoph Federrath,$^{\orcidicon{0000-0002-0706-2306}\,1,2}$
and Matthew W. Kunz$^{\orcidicon{0000-0003-1676-6126}\,3,4}$
\\
$^{1}$Research School of Astronomy and Astrophysics, Australian National University, Canberra, ACT 2611, Australia\\
$^{2}$Australian Research Council Centre of Excellence in All Sky Astrophysics (ASTRO3D), Canberra, ACT 2611, Australia\\
$^{3}$Department of Astrophysical Sciences, Princeton University, Peyton Hall, Princeton, NJ 08544, USA\\
$^{4}$Princeton Plasma Physics Laboratory, PO Box 451, Princeton, NJ 08543, USA
}

\date{Accepted XXX. Received YYY; in original form ZZZ}

\pubyear{2023}

\newcommand\Eq[1]{equation\,(\ref{#1})}
\newcommand\Fig[1]{Fig.~\ref{#1}}
\newcommand\Figure[1]{Figure~\ref{#1}}
\newcommand\Sec[1]{Sec.~\ref{#1}}
\newcommand\Tab[1]{Table~\ref{#1}}
\newcommand\App[1]{Appendix~\ref{#1}}

\newcommand{\Mach}{{\mathcal{M}}}      
\newcommand{\Pm}{{\rm Pm}}
\renewcommand{\Re}{\text{Re}} 
\newcommand{\Rm}{{\rm Rm}}

\renewcommand{\vec}[1]{\boldsymbol{#1}}	
\newcommand{\dd}{\mathrm{d}}        

\newcommand{\cm}{{\rm cm}}    
\newcommand{\pc}{{\rm pc}}     
\newcommand{\kpc}{{\rm kpc}}  

\newcommand{\G}{{\rm G}}      

\newcommand{\K}{{\rm K}}      

\newcommand{\Emag}{E_{\rm{m}}/E_{\rm{m0}}}

\newcommand{\bcdot}{\,\mathbf{\cdot}\,}
\newcommand{\grad}{\mathbf{\nabla}}
\newcommand{\btimes}{\,\mathbf{\times}\,}

\newcommand{\Esat}{E_{\rm{mag}}/E_{\rm{kin}}}
\newcommand{\ratio}{E_{\rm{mag}}/E_{\rm{kin}}}
\newcommand{\Rmcrit}{{\rm Rm}_{\rm crit}}
\newcommand{\Rk}{{\rm Re}}
\newcommand{\Gammasat}{\Gamma_{\rm sat}}
\newcommand{\expcoeff}{\alpha}
\newcommand{\Rmhyper}{{\rm Rm_{\rm h}}}
\newcommand{\magnetisation}{r_{\rm Larmor}/L_{\rm box}}
\newcommand{\initmagnetisation}{(r_{\rm Larmor}/L_{\rm box})_{\rm i}}
\newcommand{\Einit}{(E_{\rm{mag}}/E_{\rm{kin}})_{\rm{i}}}
\newcommand{\betainit}{\beta_{\rm{i}}}
\newcommand{\ngrid}{N_{\rm{grid}}}
\newcommand{\nppc}{N_{\rm{ppc}}}
\newcommand{\vel}{\mathbf{v}}
\renewcommand{\vec}[1]{\mathbf{#1}}
\newcommand{\Vturb}{V_{\rm turb}}
\newcommand{\Vtherm}{V_{\rm th}}

\graphicspath{{./}{figs/}}

\begin{document}
\label{firstpage}
\pagerange{\pageref{firstpage}--\pageref{lastpage}}
\maketitle

\begin{abstract}
The intracluster medium of galaxy clusters is an extremely hot and diffuse, nearly collisionless plasma, which hosts dynamically important magnetic fields of ${\sim}\mu \G$ strength. Seed magnetic fields of much weaker strength of astrophysical or primordial origin can be present in the intracluster medium. In collisional plasmas, which can be approximated in the magneto-hydrodynamical (MHD) limit, the turbulent dynamo mechanism can amplify weak seed fields to strong dynamical levels efficiently by converting turbulent kinetic energy into magnetic energy. However, the viability of this mechanism in weakly collisional or completely collisionless plasma is much less understood. In this study, we explore the properties of the collisionless turbulent dynamo by using three-dimensional hybrid-kinetic particle-in-cell simulations. We explore the properties of the collisionless turbulent dynamo in the kinematic regime for different values of the magnetic Reynolds number, $\Rm$, initial magnetic-to-kinetic energy ratio, $\Einit$, and initial Larmor ratio, $\initmagnetisation$, i.e., the ratio of the Larmor radius to the size of the turbulent system. We find that in the `un-magnetised' regime, $\initmagnetisation > 1$, the critical magnetic Reynolds number for the dynamo action $\Rmcrit \approx 107 \pm 3$. In the `magnetised' regime, $\initmagnetisation \lesssim 1$, we find a marginally higher $\Rmcrit = 124 \pm 8$. We find that the growth rate of the magnetic energy does not depend on the strength of the seed magnetic field when the initial magnetisation is fixed. We also study the distribution and evolution of the pressure anisotropy in the collisionless plasma and compare our results with the MHD turbulent dynamo. 
\end{abstract}

\begin{keywords}
dynamo -- turbulence --  magnetic fields -- methods: numerical -- galaxies: clusters: intracluster medium -- plasmas 
\end{keywords}



\section{Introduction}

The intracluster medium  (ICM) of galaxy clusters is an extremely hot (${\sim}10^{7}$--$10^{8}$~K) and diffuse (${\sim}10^{-2}$--$10^{-3}~{\rm cm}^{-3}$) plasma. As a result, the mean free path between Coulomb collisions in the ICM is large ($\lambda_{\rm mfp} \sim 30~{\rm kpc}$), with only limited scale separation between it and the characteristic length scales of bulk flows and temperature profiles  ($L_{\rm ICM} \sim 100~\kpc$). As a result, the ICM is said to be `weakly collisional' \citep{Simionescu+2019,Kunz+2022}. The ICM is also turbulent, with chaotic fluid motions driven by several physical processes like galaxy mergers, wakes of infall events, and feedback from active galactic nuclei \citep{Subramanian+2006,Banerjee&Sharma2014,Mohapatra&Sharma2019}. Turbulent velocities of ${\approx}160$~km/s have been observed in the Perseus cluster by \cite{HitomiCollaboration2016}, significantly smaller than the typical thermal speed of the hot ICM plasma, $\Vtherm \sim 1000$~km/s. The implied sonic Mach number of ${\sim} 0.1 - 0.35$ is low, but not atypical of the sub-sonic turbulence that is routinely inferred from high-resolution X-ray spectroscopy \citep[e.g.,][]{Sanders2011, Gatuzz+2022a, Gatuzz+2022b, Gatuzz+2023}. Merging galaxy clusters can have higher turbulent speeds and Mach numbers compared to relaxed clusters \citep{DominguezFernandez+2019}.

The evolution of magnetic fields in weakly collisional and collisionless plasmas in the presence of such turbulence has been studied with increasing interest in recent years \citep[e.g.,][]{Schekochihin+2005b,Schekochihin&Cowley2006,Mogavero&Schekochihin2014,Santos-Lima+2014,Melville+2016, Rinconetal2016,St-Onge&Kunz2018, St-Ongeetal2020,Rappaz&Schober2023}. However, detailed numerical studies investigating the amplification of magnetic fields via the collisionless turbulent dynamo remain in short supply. Such numerical experiments have significantly more computational cost when compared to their collisional MHD counterparts. To understand the collisionless turbulent dynamo, where a fluid description of the plasma is no longer suitable, one must resort to a kinetic treatment of the plasma. \cite{Rinconetal2016} performed numerical simulations of the collisionless turbulent dynamo by solving the Vlasov equation in six dimensions, demonstrating in a proof-of-concept manner that a turbulent dynamo mechanism is plausible in collisionless plasma. \cite{St-Onge&Kunz2018} used a hybrid-kinetic particle-in-cell (PIC) code to explore different regimes of the collisionless turbulent dynamo.

The magnetic Reynolds number, $\Rm$, is the ratio of the inductive motions (which amplify magnetic fields) and magnetic diffusion which decays magnetic fields. The $\Rm$ is an important parameter for dynamo action, and it has been shown that there exists a critical magnetic Reynolds number, $\Rmcrit$, above which the amplification of magnetic fields by the MHD turbulent dynamo is possible \citep{Moffatt1978}. However, the critical magnetic Reynolds number of the collisionless turbulent dynamo and how it depends on the initial conditions of the plasma have not been explored in detail by previous studies. The ICM is expected to have a high magnetic Reynolds number, $\Rm \sim 10^{27} - 10^{29}$ \citep{Schekochihin&Cowley2006}.

Below the critical value for $\Rm$, magnetic diffusion dominates and amplification of magnetic fields is not feasible \citep{Haugen+2004a, Schoberetal2012, Federrathetal2014ApJ, Seta+2020}. For the MHD turbulent dynamo driven by Kolmogorov-like (incompressible) turbulence, $\Rmcrit\sim 220$ for a plasma with magnetic Prandtl number, $\Pm \sim 1$ \citep{Seta+2020}. This is much lower than $\Rmcrit \gtrsim 1600$ obtained by \cite{Rinconetal2016} for the turbulent dynamo in a collisionless plasma. To first demonstrate and then understand this significant difference between the two regimes, we systematically study $\Rmcrit$ for the collisionless turbulent dynamo.

In this study, we explore how the growth of magnetic energy by the collisionless turbulent dynamo depends on $\Rm$ and estimate $\Rmcrit$ for the collisionless dynamo in different regimes via numerical experiments. We also study how the properties of the collisionless turbulent dynamo depend on the initial magnetic-field strength. We use a hybrid-kinetic PIC module that we have developed within the FLASH code \citep{Fryxelletal2000} to perform numerical simulations of the collisionless turbulent dynamo in the context of the ICM similar to \cite{St-Onge&Kunz2018}. This approach can also be extended to understand other collisionless and `magnetised' astrophysical plasmas, like the solar wind and the accretion flow onto the supermassive black hole at the Galactic centre.

The rest of this study is organised as follows. We discuss the hybrid-kinetic equations and how we solve these equations numerically in \Sec{section:methods-hybrid_kinetics}. We describe the components we add to the hybrid-kinetic equations to simulate the collisionless turbulent dynamo and the initial conditions of our simulations in \Sec{section:methods_collisionless-turbulent-dynamo}. We study the collisionless turbulent dynamo in the kinematic regime and measure the critical magnetic Reynolds number in \Sec{section:Rmstudy}. We discuss how the initial plasma beta affects the growth rate of the collisionless turbulent dynamo in \Sec{section:initseedstudy}. In \Sec{section:kinetic_instabilities}, we discuss kinetic instabilities that can facilitate turbulent dynamo action in a collisionless plasma and determine the distribution and evolution of the pressure anisotropy. We compare the properties of the collisionless turbulent dynamo with the well-studied MHD turbulent dynamo in \Sec{section:MHDdynamo}, and present our conclusions in \Sec{section:conclusions}.

\section{Methods: Hybrid-kinetics and collisionless turbulent dynamo}
\label{section:methods-hybrid_kinetics}
To study the turbulent dynamo in collisionless plasma, we have developed a hybrid-kinetic PIC module within the FLASH code  \citep{Fryxelletal2000}. We numerically solve the hybrid-kinetic equations on a uniform and triply periodic computational domain and use a prescribed driving to inject turbulence into the plasma.
Below we describe the equations of hybrid-kinetics and the details of our numerical implementation.

\subsection{Hybrid-kinetic equations}
In the hybrid-kinetic treatment, the positively charged ions are evolved as collisionless macro-particles and the electrons are treated as a massless, neutralizing fluid \citep{Winske+2022}. In this work, we consider protons to be the only positively charged particles of the plasma. The equations of motion for each particle in the presence of electromagnetic fields ($\vec{E}$ and $\vec{B}$) and turbulent driving ($\vec{f}$) can be written as
\begin{align}
        & \vel_{n} = \frac{\dd \mathbf{r}_{n}} {\dd t},
        \label{eqn:part_evolution2} \\
        & \frac{\dd \vel_{n} }{\dd t} = \frac{q}{m}\left(\mathbf{E}+\vel_{n} \btimes \mathbf{B}\right) + \mathbf{f},  \label{eqn:part_evolution1} 
\end{align}
where $q, m, \vec{r}_n,$ and $\vec{v}_n$ are the charge, mass, position, and velocity of the $n^{\rm th}$ macro-particle, respectively, and ${n} = 1, 2, \cdots, {N}$, where ${N}$ is the total number of macro-particles. 
The first term on the right-hand side of \Eq{eqn:part_evolution1} denotes the Lorentz force, which captures the acceleration of charged particles in electromagnetic fields, and the second term is the turbulent driving term (further described in \Sec{sec:turbdriv}). We solve \Eq{eqn:part_evolution2} and \Eq{eqn:part_evolution1} using the Boris integration scheme, which is commonly used in PIC simulations as it is designed to be energy-conserving and stable \citep{Boris1970,Kunzetal2014,Zenitani&Umeda2018}.

Next, we consider the evolution of electrons described by the Vlasov--Landau equation. Expanding the electron distribution function in powers of the mass ratio of electrons and protons, $(m_{\rm e}/m_{\rm p})^{1/2}$, we can re-write the momentum equation for the electron fluid in the form of a generalised Ohm's law  \citep{Rosin+2011},
\begin{equation}
\begin{aligned}
\mathbf{E} ={-} & \frac{1}{\rho_{\rm e}}\left(\mathbf{J}_{\mathrm e} \btimes \mathbf{B} \right) + \frac{\grad p_{\mathrm e}}{\rho_{\rm e}} + \mu_{0} \eta \mathbf{J}- \mu_{0} \eta_{\mathrm h} \nabla^{2} \mathbf{J}, \\
\end{aligned}
\label{eqn:ohms_law}
\end{equation}
where $\mathbf{J}_{\mathrm e}$, $\mathbf{J}$, $\rho_{\rm e}$ and $p_{\mathrm e}$ are the electron current, the total current, electron charge density and the electron pressure, respectively. The first term on the right-hand side of \Eq{eqn:ohms_law} is the magnetic force exerted on the electrons. The second term is the thermoelectric term, which can be responsible for generating seed magnetic fields when electron pressure and density gradients are misaligned, also known as the \cite{Biermann1950} battery term. The third term is the Ohmic dissipation arising from ion-electron collisions, where $\eta$ is the magnetic diffusivity, and is added to the Ohm's law as a sink for magnetic energy. $\mu_0$ is the magnetic permeability constant. The final term on the right-hand side is the numerical hyper-diffusivity ($\eta_{\mathrm h}$), which is an additional higher-order dissipative term. This term is primarily introduced to damp the propagation of grid-scale dispersive waves.

We assume that the electron pressure is isotropic and satisfies an isothermal equation of state, $p_{\mathrm e} \propto {\rho_{\rm e}}$. We can also assume that the plasma is quasineutral, which implies that the charge densities of ions and electrons are the same ($\rho_{\mathrm I} = -{\rho_{\rm e}}$). This assumption is valid for scales much larger than the Debye length, which is the length scale below which significant charge separation is possible in a plasma (e.g., ${\sim}10^{-13}~\pc$ for the hot ICM with a number density ${\sim}10^{-2}~\cm^{-3}$ and a temperature ${\sim}10^{7}~\K$). 

The total current can be written as $\mathbf{J} = \mathbf{J}_{\mathrm{I}} + \mathbf{J}_{\mathrm e}$, i.e., the sum of the electron current and the ion current ($\mathbf{J}_{\mathrm {I}}$). Using these, the above Ohm's law can be re-written as 
\begin{equation}
\begin{aligned}
\mathbf{E} = \frac{\left( \mathbf{J} - \mathbf{J}_{\mathrm{I} } \right) \btimes \mathbf{B}}{\rho_{\rm I}} - \frac{\grad p_{\mathrm e}}{\rho_{\rm I}}+ \mu_{0} \eta \mathbf{J}- \mu_{0} \eta_{\mathrm h} \nabla^{2} \mathbf{J}.
\end{aligned}
\label{eqn:ohms_law_hp}
\end{equation}
From Ampere's law, the total current, $\mathbf{J}$, can be written as 
\begin{equation}
\begin{aligned}
\mathbf{J} = \frac{\grad \btimes \mathbf{B}}{\mu_{0}}.
\label{eqn:amperes_law}
\end{aligned}
\end{equation}

The electromagnetic fields are evolved on a 3D computational grid, while the macro-particles represent the ions, which move in the spatial 3D computational domain and are coupled to the electric and magnetic fields through the Lorentz force. The evolution of particles and electromagnetic fields are coupled via interpolation operations to and from the grid. In particular, the quantities $\rho_{\rm I}$ and $\mathbf{J}_{\mathrm{I}}$ are the source terms in the generalised Ohm's law (\Eq{eqn:ohms_law_hp}). As the positions and velocities of the particles evolve (as described by \Eq{eqn:part_evolution2} and \Eq{eqn:part_evolution1}), the charge density and ion currents change, thereby changing the electric field. After determining the electric field, the magnetic field can be calculated from Faraday's equation, 
\begin{align}
    & \frac{\partial \mathbf{B}}{\partial t} = - \grad \btimes \mathbf{E}.
    \label{eqn:faradays_law2}
\end{align}

Finally, the updated electromagnetic fields are interpolated from the computational grid to the particles to evolve them further in time. We use the cloud-in-cell algorithm for grid-to-particle and particle-to-grid interpolations. We use the predictor-predictor-corrector algorithm developed by \citet{Kunzetal2014} to evolve the set of equations presented here.

\subsection{Corrections for interpolated electromagnetic fields}
If thermoelectric and resistive effects are not included ($\eta = \eta_{\rm h} = 0$), the electric and magnetic fields are orthogonal to each other by construction. This is true for fields calculated on the computational grid. However, when these fields are interpolated to the particles, this might no longer hold true. The generation of these un-physical spurious electric field components parallel to the magnetic field due to interpolation errors can accelerate particles and lead to unwanted numerical heating of the plasma. To correct for this anomaly, we introduce the following corrections to the interpolated electric field on particle positions \citep{Lehe2009},
\begin{equation}
    \textbf{E}_{\rm int}^{*} = \textbf{E}_{\rm int} + \left[ (\textbf{E} \bcdot \textbf{B})_{\rm int} - \textbf{E}_{\rm int} \bcdot \textbf{B}_{\rm int}  \right] \frac{\textbf{B}_{\rm int}}{B_{\rm  int}^{2}},
\end{equation}
where $\textbf{E}_{\rm int}$ and $\textbf{B}_{\rm int}$ are the electric and magnetic fields interpolated onto the particles, $(\textbf{E} \bcdot \textbf{B})_{\rm int}$ is the dot product of $\textbf{E}$ and $\textbf{B}$ on the grid interpolated to the particles, and $\textbf{E}_{\rm int}^{*}$ is the modified interpolated electric field on the particles. This correction ensures that $\textbf{E}_{\rm int}^{*} \bcdot \textbf{B}_{\rm int} = (\textbf{E} \bcdot \textbf{B})_{\rm int}$ and guarantees that the electromagnetic fields are orthogonal after interpolation from the grid to the particle positions.

\label{section:methods_collisionless-turbulent-dynamo}
\subsection{Turbulence driving} 
\label{sec:turbdriv}
We model the turbulent driving field, $\vec{f}$, in \Eq{eqn:part_evolution1} by using the Ornstein--Uhlenbeck process through TurbGen \citep{CF2010,FederrathEtAl2022ascl}. We drive turbulence on large length scales, i.e., on wave numbers satisfying $kL_{\rm box}/2\pi =(1,3)$, where $L_{\rm box}$ is the side length of our cubic computational domain. The amplitude of the driving is controlled by a parabolic function that peaks at $kL_{\rm box}/2\pi = 2$ and goes to zero at $kL_{\rm box}/2\pi = 1, 3$.

The nature of the turbulent driving affects the properties of the MHD turbulent dynamo as shown in previous works \citep{CFetal11, AchikanathEtAl2021, Seta&Federrath2022}. In this study, to maximise the efficiency of the turbulent dynamo, we focus on purely solenoidal driving $(\grad \bcdot \textbf{f} = 0)$, which injects solenoidal acceleration modes into the plasma. The magnitude of the driving amplitude controls the amount of kinetic energy being injected by the turbulence and determines the Mach number, $\mathcal{M}$, of the plasma, which is defined as
\begin{equation}
    \mathcal{M} = \frac{\Vturb}{\Vtherm},
    \label{eqn:Mach}
\end{equation}
where $\Vturb$ is the turbulent speed and $\Vtherm$ is the thermal speed of the plasma. The eddy-turnover time is defined as $t_{0} = L_{\rm box}/(2 \Vturb)$, where $ L_{\rm box}/2$ is the characteristic turbulence driving scale. We note that $\Vturb$ is the time-averaged turbulent speed calculated after steady state turbulence is established in our simulations.

\subsection{Plasma cooling}
\label{sec:plascool}
We set up subsonic turbulence with $\mathcal{M} \sim 0.25$ in our numerical simulations, comparable to the Mach number of the ICM, and drive continuously to maintain a statistically steady Mach number throughout our runs. The turbulent energy is injected primarily on large scales. This energy drives large-scale turbulent velocities, and through the turbulent cascade, the energy is transferred to smaller and smaller scales, where it ultimately dissipates \citep{Frisch1995,Federrath+2021} and heats the gas. This leads to a gradual increase in the temperature of the plasma if no cooling is applied. As a consequence, $\Vtherm$ increases, leading to a gradual decrease in the Mach number. Thus, without cooling, it is impossible to maintain statistically steady turbulence. Previous numerical studies in MHD have shown that the properties of the MHD dynamo (such as the growth rate and saturation level) are sensitive to the turbulent Mach number \citep{CFetal11, Seta&Federrath2021b, AchikanathEtAl2021}.

To mitigate the increase in $\Vtherm$ and to enable the study of the collisionless turbulent dynamo in a statistically steady state (with a statistically stationary Mach number), we implement a cooling method (Achikanath Chirakkara et al. 2023, in prep.) to remove excess thermal energy from the plasma. We implement this cooling by resetting $\Vtherm$ to its target (constant) value, on the sound-crossing timescale (${\sim}L_{\rm box } / \Vtherm$), as follows. The diagonal components of the ion pressure tensor in each cell are used along with that cell's number density to compute local thermal speeds in the $x$, $y$, and $z$ directions; these three speeds are then interpolated to each particle position to form the vector $\vec{\Vtherm^{\rm p}}$.  
The corresponding three components of each particle's velocity, measured relative to the local bulk flow velocity (interpolated from the grid to the particle position), are then rescaled using the ratio of the (constant) target thermal speed and $\Vtherm^{\rm p}=|\vec{\Vtherm^{\rm p}}|$. This keeps the direction of each particle's velocity unchanged, while reducing each particle's `random' velocity so that the plasma temperature is held at the target temperature. This procedure does not change the shape of the velocity-space distribution function, but rather decreases its standard deviation to cool the plasma and maintain isothermal conditions locally at each grid cell. As a result, the plasma also remains globally isothermal throughout the computational domain.

\subsection{Initial conditions and simulation models}
\label{sec:initcond}

\subsubsection{Numerical criteria for resolving the Larmor radius}
\label{sec:resoturb}
The main parameters of our turbulent dynamo experiments are the Mach number ($\mathcal{M}$) and the initial ratio of the magnetic energy to the turbulent kinetic energy ($\Einit$). For collisionless plasmas, there is an additional parameter -- the initial Larmor radius of the particles. We quantify the level of magnetisation of the plasma using the Larmor ratio defined as the ratio of the Larmor radius ($r_{\rm Larmor}$) to the box length ($L_{\rm box}$),
\begin{equation}
    \frac{r_{\rm Larmor}}{L_{\rm box}} = 
    \frac{m \Vtherm^{\rm p}}{q B L_{\rm box}},
    \label{eqn:Larmor_ratio_def}
\end{equation}
where $\Vtherm^{\rm p}$ is the thermal speed interpolated to the particle position and $B$ is the magnetic-field strength. Because the probability density function of the magnetic-field strength in the kinematic regime of the MHD turbulent dynamo follows a lognormal distribution \citep{Seta&Federrath2021b}, we use the average of the logarithmic value of the Larmor ratio calculated from all the particles to quantify the mean Larmor radius in our numerical simulations. The initial Larmor ratio can be written as $\initmagnetisation = {m \Vtherm^{\rm p} }/{(q B_{\rm i} L_{\rm box})}$, where $B_{\rm i}$ is the initial magnetic-field strength. 
The ratio of magnetic energy to kinetic energy can be written as
\begin{equation}
     \frac{E_{ {\rm mag} }}{E_{ {\rm kin} }} = \frac{B^{2}/(2\mu_{0})}{\rho_{\rm m} \Vturb^{2}/2} = \frac{ B^{2}}{\mu_{0} \rho_{\rm m}  \Vturb^{2}},
    \label{eqn:ratio_def}
\end{equation}
where $\mu_{0}$ is the vacuum permeability and $\rho_{\rm m}$ is the mass density of ions. Further, the initial magnetic to kinetic energy ratio can be written as, $\Einit = { B_{\rm i}^{2}}/{(\mu_{0} \rho_{\rm m}  \Vturb^{2})} $.
As magnetic energy grows due to dynamo action, the Larmor ratio decreases proportionally to the increase in the magnetic-field strength. To ensure that we resolve the average particle Larmor radius throughout our simulations, up to the saturation stage of the dynamo, we impose the following constraint on the initial conditions. 

$\Esat$ is the ratio of the magnetic energy to the kinetic energy of the dynamo and quantifies the efficiency by which the dynamo converts turbulent kinetic energy to magnetic energy. Assuming a maximum possible level for this ratio at the saturation stage of the dynamo, $(\Esat)_{\rm sat} = 1$~\citep[it is usually $< 1$; see][for the MHD dynamo]{CFetal11,Federrathetal2014ApJ,AchikanathEtAl2021}, from \Eq{eqn:ratio_def} we have
\begin{equation}
     \frac{B_{\rm sat}^{2}}{2\mu_{0}} = \frac{1}{2}{\rho_{\rm m} }\Vturb^{2},
     \label{eqn:bsat}
\end{equation}
or the magnetic-field strength at saturation $B_{\rm sat} = (\rho_{\rm m} \mu_{0})^{1/2}  \Vturb$. The Larmor ratio at saturation can therefore be written as
\begin{equation}
     \left( \frac{r_{\rm Larmor}}{L_{\rm box}} \right)_{\rm sat} = \frac{m \Vtherm^{\rm p} }{q B_{\rm sat} L_{\rm box}} > \frac{1}{\ngrid},
     \label{eqn:rlbyL}
\end{equation}
where $\ngrid$ is the number of grid points along a linear dimension of the simulation cube. Using the expression for $B_{\rm sat}$, \Eq{eqn:rlbyL} can be simplified to
\begin{equation}
    \frac{m  \ngrid }{q \mu_{0}^{1/2} \mathcal{M} } > \rho_{\rm m}^{1/2} L_{\rm box}.
    \label{eqn:rlbyLcond}
\end{equation}
Further from \Eq{eqn:Larmor_ratio_def} and \Eq{eqn:ratio_def}, we can write
\begin{equation}
    \rho_{\rm m}^{1/2} L_{\rm box} \propto   \left[{\Einit}^{1/2} \, \initmagnetisation\right]^{-1},
    \label{eqn:rlbyLcondpropto}
\end{equation}
where the proportionality constant is $m /{(q \mu_{0}^{1/2} \mathcal{M})} $.
From the above expressions, we obtain the following constraint that links the initial conditions with the grid resolution
\begin{equation}
   \Big(\frac{E_{\rm mag}}{E_{\rm kin}}\Big)_{\rm i}^{1/2} \Big( \frac{r_{\rm Larmor}}{L_{\rm box}} \Big)_{\rm i}  > \frac{1}{\ngrid}.
   \label{eqn:res_criteria}
\end{equation}

Thus, given a value for the grid resolution, $\ngrid$, the criterion derived in \Eq{eqn:res_criteria} limits the range of magnetisation regimes and the initial ratio of magnetic to turbulent energies that we can explore through our simulations.

\subsubsection{Resistivity and hyper-resistivity}
\label{sec:resihype}
We define the magnetic Reynolds number ($\Rm$) as 
\begin{equation}
    \Rm = \frac{\Vturb (L_{\rm box }/2)}{\eta},
\end{equation}
where $L_{\rm{box}}/2$ is the turbulent driving scale (see Sec.~\ref{sec:turbdriv}) and $\eta$ is the Ohmic diffusivity.  Similarly, we can define the kinetic Reynolds number ($\Rk$) using the viscosity in place of the Ohmic diffusivity. However, the viscosity is set by wave-particle interactions in the collisionless plasma and, unlike the magnetic diffusivity, it is not a parameter we can control in our simulations. The magnetic Prandtl number ($\Pm$), defined as the ratio of the magnetic Reynolds number to the kinetic Reynolds number, is therefore {\em a priori} unknown.

We define the hyper-resistive Reynolds number ($\Rmhyper$) as 
\begin{equation}
    \Rmhyper = \frac{V_{\rm{turb}} (L_{\rm{box}}/2)^{3}}{ \eta_{\rm h}},
\end{equation}
where $\eta_{\rm h}$ is the hyper-resistivity coefficient, as in \Eq{eqn:ohms_law_hp}. 

We choose the values for the resistivity and hyper-resistivity such that the dissipation due to these terms is greater than the corresponding numerical dissipation. To estimate the numerical dissipation, we assume a Kolmogorov spectrum for the velocity field from the driving scale down to half a grid cell spacing and derive that the magnetic Reynolds number, which can be resolved well at a given grid resolution, $\ngrid$, scales as $\Rm \sim {\rm A} \ngrid^{4/3}$, where the value of the coefficient was estimated from MHD simulations to be ${\rm A} = 0.5 - 2$ \citep{Federrath+2011,McKee+2020}. We set the hyper-resistivity similarly, assuming that the appropriately resolvable hyper-resistive Reynolds number scales as $\Rmhyper \sim  {\rm A}_{\rm h} \ngrid^{10/3}$. Given a value of $\ngrid$ and assuming ${\rm A} = {\rm A}_{\rm h} = 1$, these expressions constrain the range of $\Rm$ and $\Rmhyper$ that we can explore. We find that for $\ngrid=120$, the maximum $\Rm$ and $\Rmhyper$ are ${\sim} 600$ and ${\sim} 8.5 \times 10^{6}$, respectively.

\subsubsection{Simulation parameters}
\label{sec:setup&table}
Our subsonic collisionless turbulent dynamo simulations use a triply periodic uniform computational domain with $\ngrid^3 = 120^3$~grid cells and $\nppc=100$~particles per cell (ppc). In order to test numerical convergence, we also perform a subset of our simulations with two other particle ($\nppc = 50$ and 200) and grid ($\ngrid = 60$ and 180) resolutions. We find our results show convergence with both types of resolutions (see \App{Appendix:res_test} for further details).

We model the magnetic seed field using a parabolic function on large scales, $k/(2\pi/L_{\rm box})=[1\dots3]$, with the maximum magnetic energy at $k = 2\pi/ (L_{\rm box}/2)$. This is identical to how we construct the turbulence driving acceleration field, using TurbGen \citep{FederrathEtAl2022ascl}, except that for the magnetic field, we only generate the field once, to be used as an initial condition. We explore a range of magnetic Reynolds numbers in our numerical experiments, $\Rm \sim 30, 60, 120, 240, 480$ and 960, up to the resolvable $\Rm$ limit with $\ngrid = 120$, and fix the hyper-resistive Reynolds number, $\Rmhyper = 8.5 \times 10^{6}$ for all our simulations. We note that $\Rm = 960$ can be marginally unresolved for $\ngrid = 120$.

For all our simulations, we tabulate the initial conditions, grid and particle resolution, measured value of the magnetic Reynolds number, Mach number, and growth rate of the collisionless turbulent dynamo in \Tab{table:sims}. We keep the thermal speed, $\Vtherm$, fixed across all our simulations. The target Mach number of the plasma determines the turbulent speed, $\Vturb$. To modify the initial Larmor ratio, $\initmagnetisation$, while maintaining a constant ratio of initial magnetic energy to kinetic energy, $\Einit$, we adjust the initial magnetic-field strength, $B_{\rm i}$. This changes both $\initmagnetisation$ and $\Einit$. We then modify the density to ensure that $\Einit$ remains the same as its previous value. To change $\Einit$ while keeping $\initmagnetisation$ constant, we solely vary the number density of the plasma. To vary the $\Rm$ of the plasma while fixing the Mach number, $\Einit$ and $\initmagnetisation$, we change the the Ohmic diffusivity, $\eta$.

\begin{table*}
\centering
\caption{List of simulations with the corresponding model name, grid resolution ($\ngrid^{3}$), particles per grid cell ($\nppc$), Mach number ($\mathcal{M}$), initial magnetic to kinetic energy ratio ($\Einit$), initial plasma beta ($\betainit$), initial Larmor ratio ($\initmagnetisation$), magnetic Reynolds number ($\Rm$) and the growth/decay rate of magnetic energy ($\Gamma$). }
\begin{tabular}{lcccccccccc}
\hline
 Ser. No. & Model & $\ngrid^{3}$ & $\nppc$ & $\mathcal{M}$ & $\Einit$ & $\betainit$ & $\initmagnetisation$ & $\Rm$ &  $\Gamma$ ($t_{0}^{-1}$)\\
\hline
 1 & \texttt{ Rm30rL1e3 } & $120^{3}$ & 100 & 0.23$\pm$0.02 & $10^{-8}$ & $10^{9}$ & $10^{3}$ & 29$\pm$2 & $-1.05\pm0.19$ \\
 2 & \texttt{ Rm60rL1e3 } & $120^{3}$ & 100 & 0.23$\pm$0.02 & $10^{-8}$ & $10^{9}$ & $10^{3}$ & 61$\pm$5 & $-0.26\pm0.04$ \\
 3 & \texttt{ Rm120rL1e3 } & $120^{3}$ & 100 & 0.23$\pm$0.02 & $10^{-8}$ & $10^{9}$ & $10^{3}$ & 121$\pm$9 & $\phantom{-}0.06\pm0.03$ \\
 4 & \texttt{ Rm240rL1e3 } & $120^{3}$ & 100 & 0.23$\pm$0.02 & $10^{-8}$ & $10^{9}$ & $10^{3}$ & 243$\pm$19 & $\phantom{-}0.25\pm0.03$ \\
 5 & \texttt{ Rm480rL1e3 } & $120^{3}$ & 100 & 0.24$\pm$0.02 & $10^{-8}$ & $10^{9}$ & $10^{3}$ & 482$\pm$37 & $\phantom{-}0.40\pm0.03$ \\
 6 & \texttt{ Rm950rL1e3 } & $120^{3}$ & 100 & 0.23$\pm$0.02 & $10^{-8}$ & $10^{9}$ & $10^{3}$ & 952$\pm$74 & $\phantom{-}0.52\pm0.05$ \\
 7 & \texttt{ Rm30rL1e2 } & $120^{3}$ & 100 & 0.23$\pm$0.02 & $10^{-8}$ & $10^{9}$ & $10^{2}$ & 29$\pm$2 & $-1.05\pm0.19$ \\
 8 & \texttt{ Rm60rL1e2 } & $120^{3}$ & 100 & 0.23$\pm$0.02 & $10^{-8}$ & $10^{9}$ & $10^{2}$ & 61$\pm$5 & $-0.25\pm0.04$ \\
 9 & \texttt{ Rm120rL1e2 } & $120^{3}$ & 100 & 0.23$\pm$0.02 & $10^{-8}$ & $10^{9}$ & $10^{2}$ & 121$\pm$9 & $\phantom{-}0.06\pm0.03$ \\
 10 & \texttt{ Rm240rL1e2 } & $120^{3}$ & 100 & 0.24$\pm$0.02 & $10^{-8}$ & $10^{9}$ & $10^{2}$ & 244$\pm$19 & $\phantom{-}0.25\pm0.03$ \\
 11 & \texttt{ Rm480rL1e2 } & $120^{3}$ & 100 & 0.24$\pm$0.02 & $10^{-8}$ & $10^{9}$ & $10^{2}$ & 482$\pm$39 & $\phantom{-}0.39\pm0.06$ \\
 12 & \texttt{ Rm960rL1e2 } & $120^{3}$ & 100 & 0.24$\pm$0.02 & $10^{-8}$ & $10^{9}$ & $10^{2}$ & 963$\pm$81 & $\phantom{-}0.44\pm0.07$ \\
 13 & \texttt{ Rm30rL10 } & $120^{3}$ & 100 & 0.23$\pm$0.02 & $10^{-8}$ & $10^{9}$ & $10$ & 29$\pm$2 & $-1.05\pm0.19$ \\
 14 & \texttt{ Rm60rL10 } & $120^{3}$ & 100 & 0.23$\pm$0.02 & $10^{-8}$ & $10^{9}$ & $10$ & 61$\pm$5 & $-0.26\pm0.04$ \\
 15 & \texttt{ Rm120rL10 } & $120^{3}$ & 100 & 0.23$\pm$0.02 & $10^{-8}$ & $10^{9}$ & $10$ & 121$\pm$9 & $\phantom{-}0.04\pm0.03$ \\
 16 & \texttt{ Rm250rL10 } & $120^{3}$ & 100 & 0.24$\pm$0.02 & $10^{-8}$ & $10^{9}$ & $10$ & 248$\pm$21 & $\phantom{-}0.15\pm0.03$ \\
 17 & \texttt{ Rm480rL10 } & $120^{3}$ & 100 & 0.24$\pm$0.02 & $10^{-8}$ & $10^{9}$ & $10$ & 483$\pm$46 & $\phantom{-}0.32\pm0.10$ \\
 18 & \texttt{ Rm970rL10 } & $120^{3}$ & 100 & 0.24$\pm$0.02 & $10^{-8}$ & $10^{9}$ & $10$ & 970$\pm$95 & $\phantom{-}0.33\pm0.08$ \\
 19 & \texttt{ Rm30rL1 } & $120^{3}$ & 100 & 0.23$\pm$0.02 & $10^{-8}$ & $10^{9}$ & $1$ & 29$\pm$2 & $-1.04\pm0.11$ \\
 20 & \texttt{ Rm60rL1 } & $120^{3}$ & 100 & 0.23$\pm$0.02 & $10^{-8}$ & $10^{9}$ & $1$ & 61$\pm$5 & $-0.26\pm0.04$ \\
 21 & \texttt{ Rm120rL1 } & $120^{3}$ & 100 & 0.24$\pm$0.02 & $10^{-8}$ & $10^{9}$ & $1$ & 124$\pm$9 & $-0.01\pm0.03$ \\
 22 & \texttt{ Rm250rL1 } & $120^{3}$ & 100 & 0.24$\pm$0.02 & $10^{-8}$ & $10^{9}$ & $1$ & 248$\pm$19 & $\phantom{-}0.10\pm0.16$ \\
 23 & \texttt{ Rm510rL1 } & $120^{3}$ & 100 & 0.25$\pm$0.02 & $10^{-8}$ & $10^{9}$ & $1$ & 514$\pm$40 & $\phantom{-}0.21\pm0.10$ \\
 \hline
 24 & \texttt{ Rm480rL1e2E1e-6 } & $120^{3}$ & 100 & 0.24$\pm$0.02 & $10^{-6}$ & $10^{7}$ & $10^{2}$ & 483$\pm$39 & $\phantom{-}0.39\pm0.06$ \\
 25 & \texttt{ Rm480rL1e2E1e-10 } & $120^{3}$ & 100 & 0.24$\pm$0.02 & $10^{-10}$ & $10^{11}$ & $10^{2}$ & 482$\pm$39 & $\phantom{-}0.39\pm0.06$ \\
 \hline
 26 & \texttt{ Rm490rL1e2RSII } & $120^{3}$ & 100 & 0.24$\pm$0.02 & $10^{-8}$ & $10^{9}$ & $10^{2}$ & 482$\pm$39 & $\phantom{-}0.39\pm0.06$ \\
 27 & \texttt{ Rm500rL1e2RSIII } & $120^{3}$ & 100 & 0.24$\pm$0.02 & $10^{-8}$ & $10^{9}$ & $10^{2}$ & 493$\pm$42 & $\phantom{-}0.36\pm0.04$ \\
 \hline
 28 & \texttt{ Rm480rL1e2$\ngrid$60 } & $60^{3}$ & 100 & 0.23$\pm$0.02 & $10^{-8}$ & $10^{9}$ & $10^{2}$ & 476$\pm$37 & $\phantom{-}0.38\pm0.06$ \\
 29 & \texttt{ Rm480rL1e2$\ngrid$180 } & $180^{3}$ & 100 & 0.24$\pm$0.02 & $10^{-8}$ & $10^{9}$ & $10^{2}$ & 482$\pm$39 & $\phantom{-}0.39\pm0.06$ \\
 \hline
 30 & \texttt{ Rm480rL1e2$\nppc$50 } & $120^{3}$ & 50 & 0.24$\pm$0.02 & $10^{-8}$ & $10^{9}$ & $10^{2}$ & 481$\pm$38 & $\phantom{-}0.38\pm0.06$ \\
 31 & \texttt{ Rm480rL1e2$\nppc$200 } & $120^{3}$ & 200 & 0.24$\pm$0.02 & $10^{-8}$ & $10^{9}$ & $10^{2}$ & 482$\pm$39 & $\phantom{-}0.39\pm0.06$ \\
 \hline
\end{tabular}
\label{table:sims}
\end{table*}

\section{Critical magnetic Reynolds number of the collisionless turbulent dynamo}
\label{section:Rmstudy}

In this section, we explore the effect of the magnetic Reynolds number on the properties of the turbulent collisionless dynamo. For this study, we fix the initial magnetic to kinetic energy ratio, $\Einit = 10^{-8}$, and vary the magnetic Reynolds number of the plasma. In addition, we determine how magnetisation affects the growth rate of the collisionless turbulent dynamo, by varying the initial Larmor ratio of the plasma, $\initmagnetisation = 10^{3}, 10^{2}, 10 $ and $1$, for each value of $\Rm$. 

\begin{figure*}
        \centering
        \begin{subfigure}[b]{0.498\textwidth}
            \centering
            \includegraphics[width=\textwidth, trim={0.15cm 0.25cm 0.3cm 0.4cm},clip]{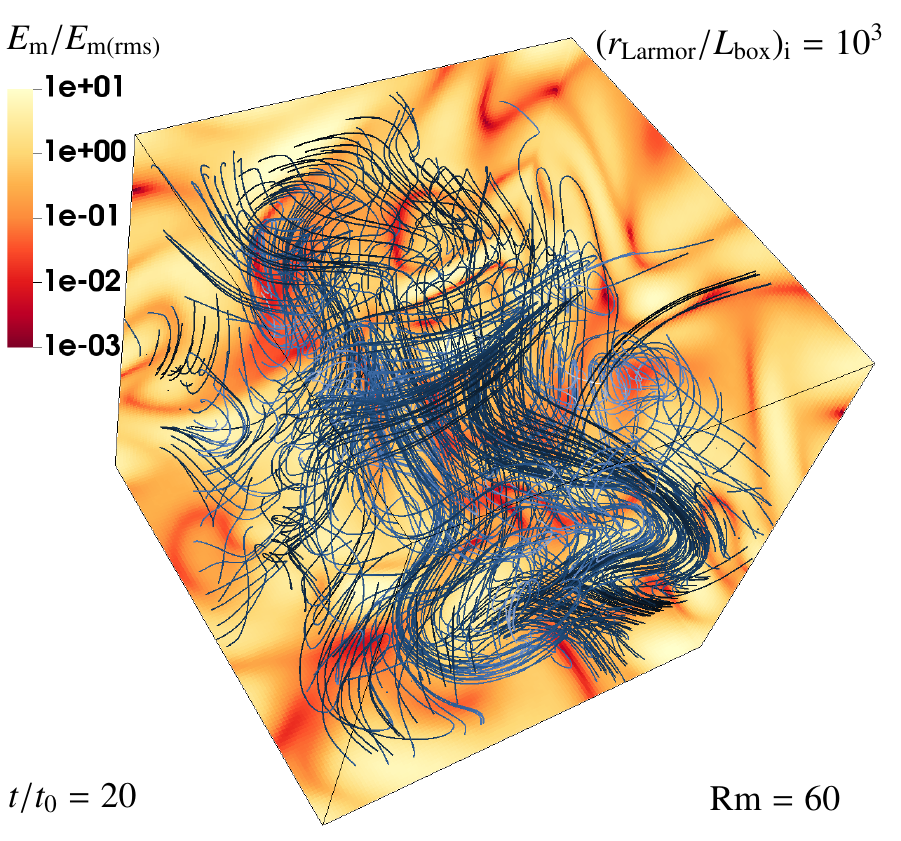}
            \label{fig:rL_1000_Rm60}
        \end{subfigure}
        \hfill
        \begin{subfigure}[b]{0.498\textwidth}  
            \centering 
            \includegraphics[width=\textwidth, trim={0.15cm 0.25cm 0.3cm 0.4cm},clip]{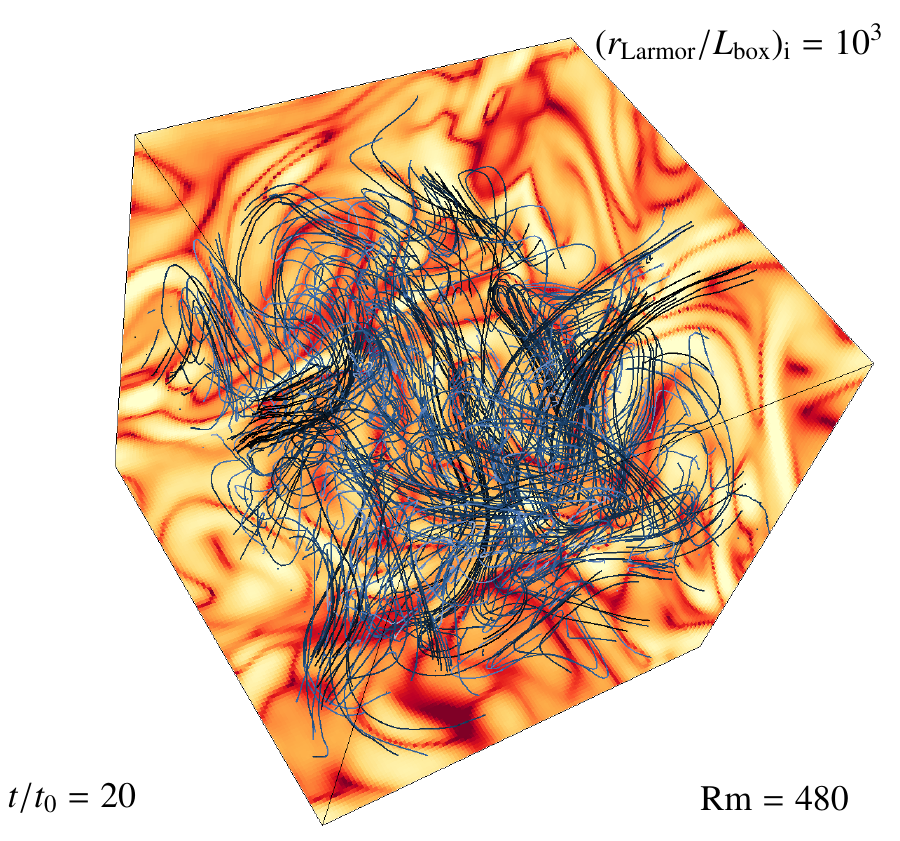}
            \label{fig:rL_1000_Rm480}
        \end{subfigure}
        \vskip\baselineskip
        \begin{subfigure}[b]{0.498\textwidth}   
            \centering 
            \includegraphics[width=\textwidth, trim={0.15cm 0.25cm 0.3cm 0.4cm},clip]{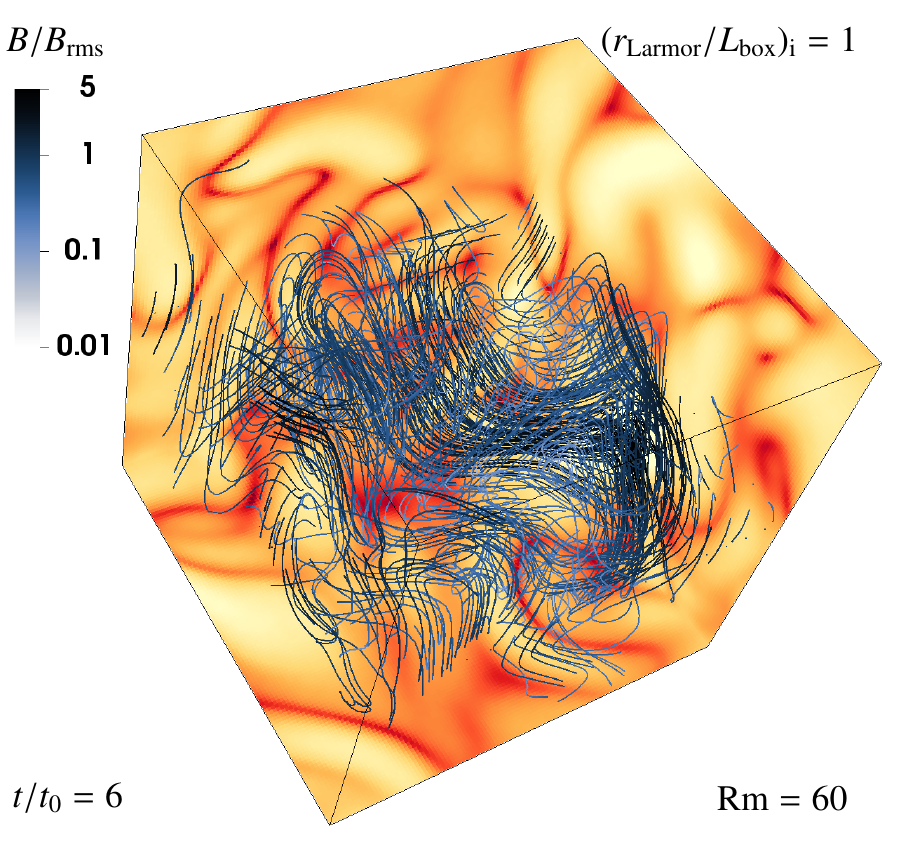}
            \label{fig:rL_1_Rm60}
        \end{subfigure}
        \hfill
        \begin{subfigure}[b]{0.498\textwidth}   
            \centering 
            \includegraphics[width=\textwidth, trim={0.15cm 0.25cm 0.3cm 0.4cm},clip]{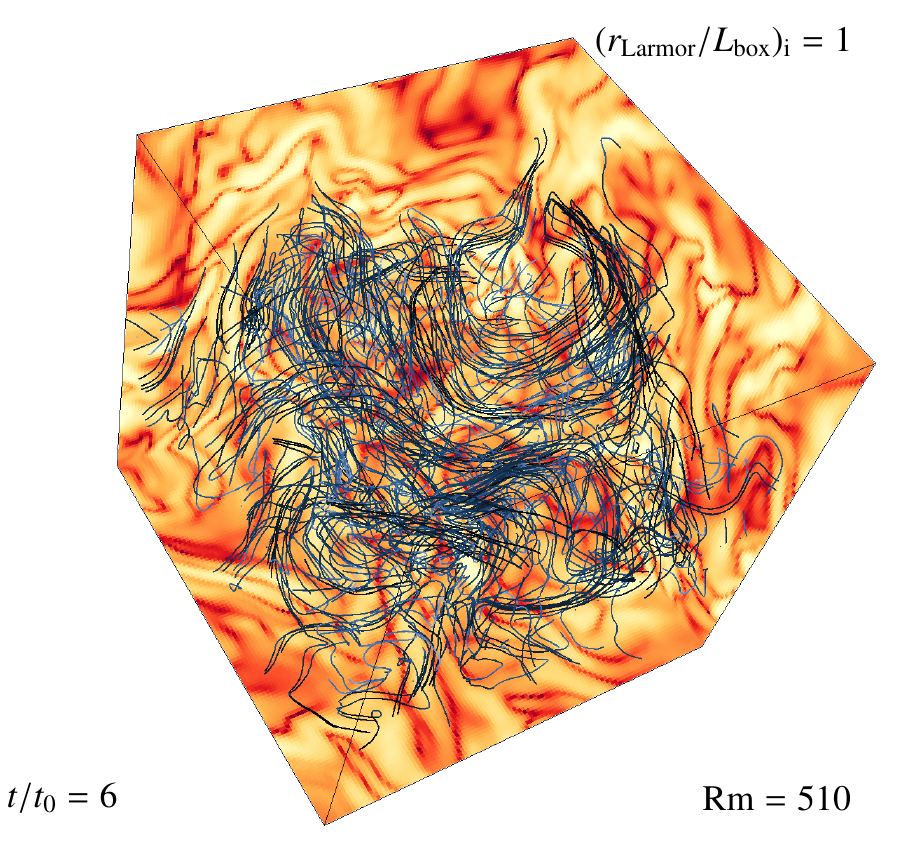}
            \label{fig:rL_1_Rm480}
        \end{subfigure}
        \caption{Slices of magnetic energy normalised to the root mean square value  ($E_{\rm m}/E_{\rm m (rms)}$) with magnetic field streamlines coloured with magnetic-field strength normalised to the root mean square value ($B/B_{\rm rms}$) in the kinematic regime of the collisionless turbulent dynamo. The top panel shows the initially `un-magnetised' simulation model, $\initmagnetisation = 10^{3}$,  for decaying (top left, $\Rm = 61 \pm 5$) and growing (top right, $\Rm = 482 \pm 37$) magnetic fields. The bottom panel shows the initially `magnetised' simulation model, $\initmagnetisation = 1$, for decaying (bottom left, $\Rm = 61 \pm 5$) and growing (bottom right, $\Rm = 514 \pm 40$) magnetic fields.}
        \label{fig:b/brms}
\end{figure*}

In \Fig{fig:b/brms}, we plot the magnetic energy normalised to its root-mean-square value ($E_{\rm m}/E_{\rm m (rms)}$; colour) along with magnetic field streamlines in the interior of the computational box coloured with the magnetic-field strength normalised to the root mean square value. The top panels show two initially `un-magnetised' simulations with $\initmagnetisation = 10^{3}$; the bottom panels show two initially `magnetised' simulations with $\initmagnetisation = 1$. The left panels show simulations with $\Rm = 60$ (decaying magnetic fields) and the right panels show simulations with $\Rm \sim 500$ (growing magnetic fields). When $\Rm\sim 500$, the magnetic energy has more small-scale structure due to the dynamo action. We also see that the topology and strength of magnetic fields vary locally. Therefore, the Larmor ratio or `magnetisation' can be very different from one spatial region to another.

\begin{figure}
\includegraphics[scale = 0.225]{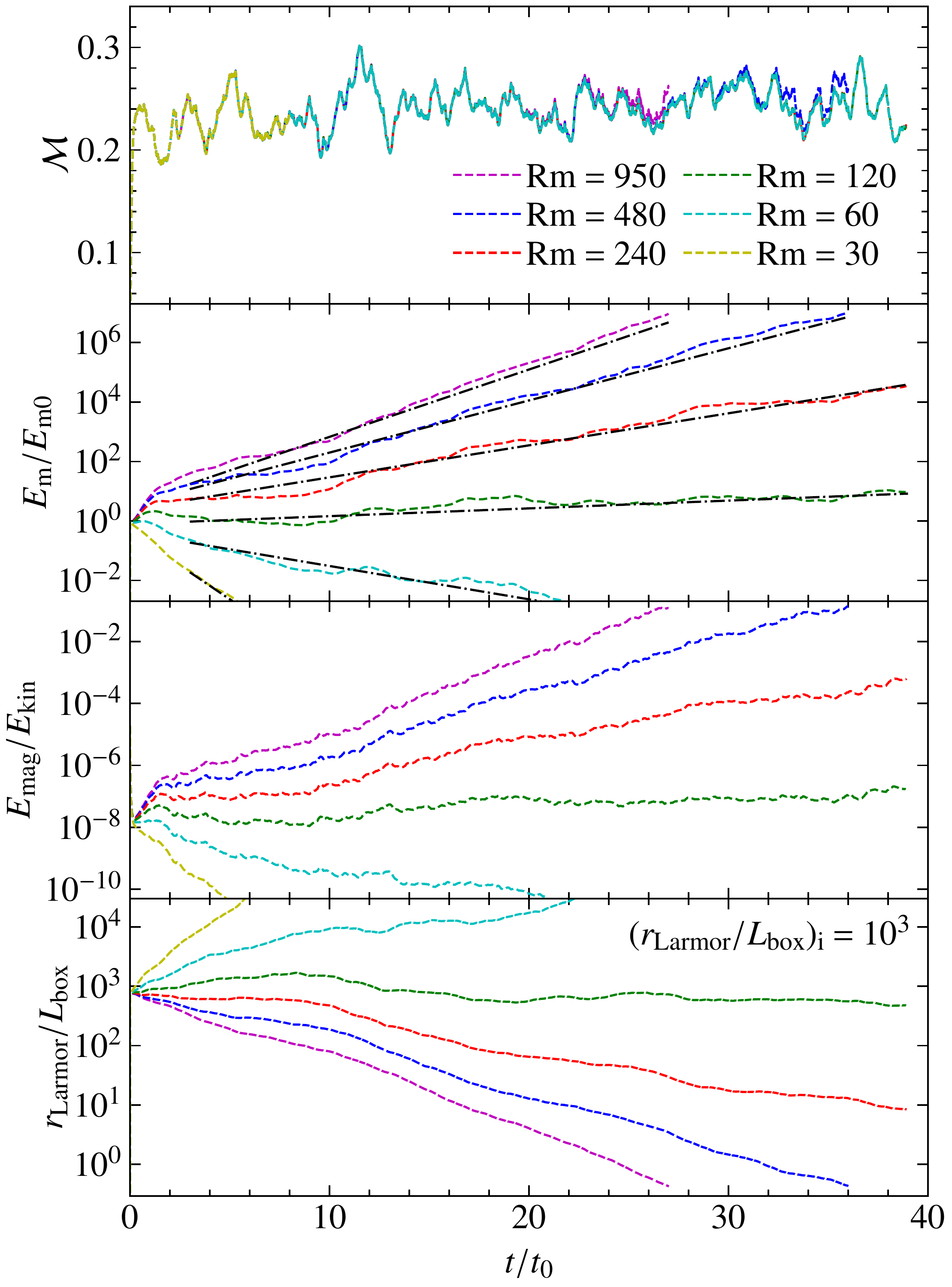}
    \caption{Mach number (first panel), magnetic energy, $\Emag$ (second panel), ratio of magnetic energy to kinetic energy, $\ratio$ (third panel) and Larmor ratio, $\magnetisation$ (fourth panel), as a function of time normalised to the turbulent eddy turnover time ($t_{0}$) for the collisionless turbulent dynamo simulations on $120^{3}$ grid cells with 100~particles per cell, for different magnetic Reynolds number ($\Rm$); see \Tab{table:sims}. For these simulations, we fix the initial Larmor ratio, $\initmagnetisation = 10^{3}$, and the initial ratio of magnetic to kinetic energy, $\Einit = 10^{-8}$. We maintain the Mach number (first panel) throughout each simulation by continuous driving and cooling (see \Sec{sec:turbdriv} and \Sec{sec:plascool}). We observe exponential growth of the magnetic energy (panels 2 and 3) for our simulations with sufficiently high $\Rm$. However, below a critical value for the magnetic Reynolds number ($\Rmcrit$), we find the magnetic energy struggles to grow and eventually decays. The fits for the growth/decay rate of magnetic energy are shown by the black dashed lines in the second panel. Finally, the fourth panel shows the evolution Larmor ratio. In the growing field cases, the Larmor radius decreases (see \Eq{eqn:Larmor_ratio_def}) and is eventually not sufficiently resolved anymore (see \Eq{eqn:res_criteria}). 
    For the present simulations, we do not follow the evolution further when the Larmor ratio drops below $\sim 0.3$ to ensure we adequately resolve the Larmor gyration of ions and to maintain a steady state Mach number across simulation models (as the amount of injected energy accepted in the form of bulk flows depends on the magnetization level of the collisionless plasma).}
\label{fig:Rmstudy_4panelplot_rL1000}
\end{figure}

\Figure{fig:Rmstudy_4panelplot_rL1000} depicts the time evolution (time normalised to the large-scale eddy turn-over time, $t_{0}$) of the dynamo simulations with $\initmagnetisation = 10^{3}$ for $\Rm = 30,\, 60,\, 120,\, 240,\, 480$ and $950$. The four panels (from first to fourth) show the evolution of the Mach number ($\mathcal{M}$), the magnetic energy normalised to the initial magnetic energy ($\Emag$), the ratio of magnetic energy to kinetic energy ($\Esat$), and the Larmor ratio ($\magnetisation$). All these quantities are averaged over the cubic computational domain.

During the initial phase of the dynamo up to $3 t_{0}$, the turbulence develops and reaches a statistically steady state. Following this initial phase, a statistically steady turbulent speed is established in the plasma and this leads to an exponential growth of magnetic energy as shown in the second panel of \Fig{fig:Rmstudy_4panelplot_rL1000}. This exponential growth phase is called the kinematic regime of the turbulent dynamo, which is the primary focus of this study. We note that in this case, the plasma is `un-magnetised' initially ($\initmagnetisation \gg 1$), and as the magnetic field grows via the dynamo mechanism, the Larmor radius of the particles decreases, eventually magnetising the plasma ($\magnetisation \lesssim 1$, as shown in the fourth panel). We measure the growth rate of the magnetic energy, $\Gamma$, by fitting an exponential curve to the magnetic energy, $\Emag = {\rm e}^{\Gamma t/t_{0}}$, in the kinematic regime. The interval over which this fit is performed is non-standardised and subject to choice, which can introduce systematic errors in the growth rate measurement. To mitigate this, we perform a systematic study on how the relative error in the growth rate depends on the fit interval chosen. This is presented in \App{Appendix:growthrateerr} and describes how we determine the errors in the estimated growth rate.

Before the exponential growth phase begins, we find a rapid initial growth in the magnetic energy until $t \lesssim 3 t_{0}$ for our simulations with high $\Rm$. We attribute this rapid growth to the turbulent generation of pressure anisotropy, which can drive kinetic instabilities. In a low $\Einit$ or high $\beta_{\rm i}$ plasma, these instabilities can be excited easily if the plasma is `magnetised' (see \Sec{section:kinetic_instabilities}). Although the plasma is initially `un-magnetised', there can be local regions where $\magnetisation < 1$, allowing the mirror and firehose instabilities to rapidly grow magnetic fields. This finding is consistent with \citet{St-Onge&Kunz2018}, who also find a rapid initial phase of magnetic energy growth in their numerical simulations for $t \lesssim 5 t_{0}$.

For higher values of $\Rm$, we see exponential amplification of the magnetic energy by the dynamo in the second and third panels of \Fig{fig:Rmstudy_4panelplot_rL1000}. However, as $\Rm$ decreases, the growth of magnetic energy by the collisionless turbulent dynamo dwindles and eventually, the magnetic energy decays for simulations with low $\Rm$. We show the fitted curves measuring the growth or decay rates of the magnetic energy as black dashed lines in the second panel of \Fig{fig:Rmstudy_4panelplot_rL1000}.

The third panel of \Fig{fig:Rmstudy_4panelplot_rL1000} shows the evolution of the ratio of magnetic energy to kinetic energy in the growth phase of the collisionless turbulent dynamo. $\ratio$ grows with time as the magnetic field is amplified by the dynamo. Eventually, we reach a regime in which the magnetic energy is comparable to the kinetic energy of the turbulent eddies at the viscous scale, where the exponential growth turns into a linear growth regime, finally leading to the saturation regime of the dynamo \citep{Seta&Federrath2020}.  

The range in which we can study the exponential growth of the collisionless turbulent dynamo becomes limited as we decrease $\initmagnetisation$, because of the requirement that we resolve the ion Larmor radius (see \Sec{sec:resoturb}). As the magnetic field grows, the average ion Larmor radius decreases, as can be seen in the fourth panel of  \Fig{fig:Rmstudy_4panelplot_rL1000}. We run all our simulations up to a magnetisation level, $\magnetisation \sim 0.3$, which ensures both that the Larmor radii of particles are resolved throughout the kinematic regime and that the Mach number is steady across simulation models (since the magnetization level affects the amount of injected energy accepted by the plasma in the form of bulk flows; see \citealt{St-Ongethesis2019}). We perform the same experiment changing the value of the magnetic Reynolds number with a different initial Larmor ratio, $\initmagnetisation = 10^{2}, 10$ and 1 and present these simulations in \App{Appendix:Larmor_ratio}. We also report the measured parameters from these simulations in \Tab{table:sims}.

\begin{figure}
    \includegraphics[scale = 0.225]{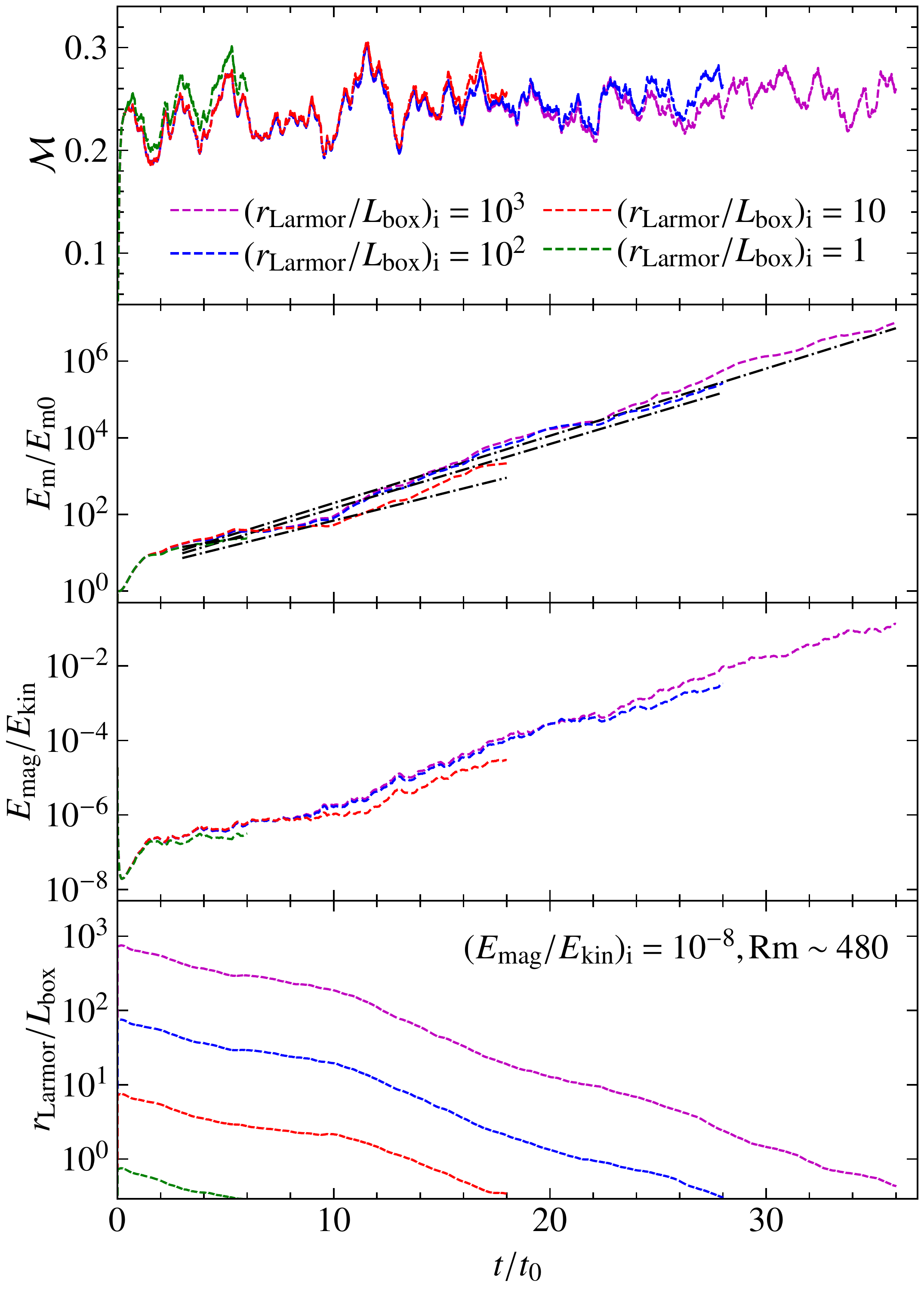}
    \caption{Same as \Fig{fig:Rmstudy_4panelplot_rL1000}, but for fixed magnetic Reynolds number, $\Rm \sim 480$, initial magnetic energy to kinetic energy ratio, $\Einit = 10^{-8}$, and varying initial Larmor ratio $\initmagnetisation = 10^{3}, 10^{2}, 10$ and 1 (see \Tab{table:sims}). We find that the growth rate of the collisionless turbulent dynamo decreases marginally with the initial Larmor ratio (see \Fig{fig:Rmtests_Growthrate}) }
    \label{fig:Rmstudy_4panelplot_Larmorratio}
\end{figure}

In \Fig{fig:Rmstudy_4panelplot_Larmorratio}, we plot the evolution of the collisionless turbulent dynamo for simulations with varying initial Larmor ratio, $\initmagnetisation$, for fixed $\Rm \sim 480$ and $\Einit = 10^{-8}$. As we decrease the $\initmagnetisation$, the growth rate of the dynamo decreases marginally. 

All our simulation models have $\mathcal{M} \sim 0.25$ maintained throughout the numerical simulation by continuous turbulent driving. We note that the value of the random seed picked to generate the turbulent driving field also influences the fine details of the evolution of the Mach number and by extension the magnetic energy. We test this for three random seed values for the simulation model with $\Rm = 480$, $\Einit = 10^{-8}$ and $\initmagnetisation = 10^{2}$, and report our findings in \App{Appendix:randomseed}. We find that local features in Mach number and magnetic energy growth are sensitive to the random seed of the turbulence driving. However, averaged over a long time in the kinematic regime, the growth rates are similar for the simulation models with different seed values (see \Tab{table:sims}). We also study the time-averaged magnetic power spectra for our simulations (see \App{Appendix:mags_spectra}) and find that on large scales the power spectra are visually consistent with the characteristic $k^{3/2}$ scaling of the MHD dynamo 
 \citep{Kazantsev1968}.

\subsection{Measuring the critical magnetic Reynolds number}

\begin{figure*}
    \includegraphics[scale = 0.25]{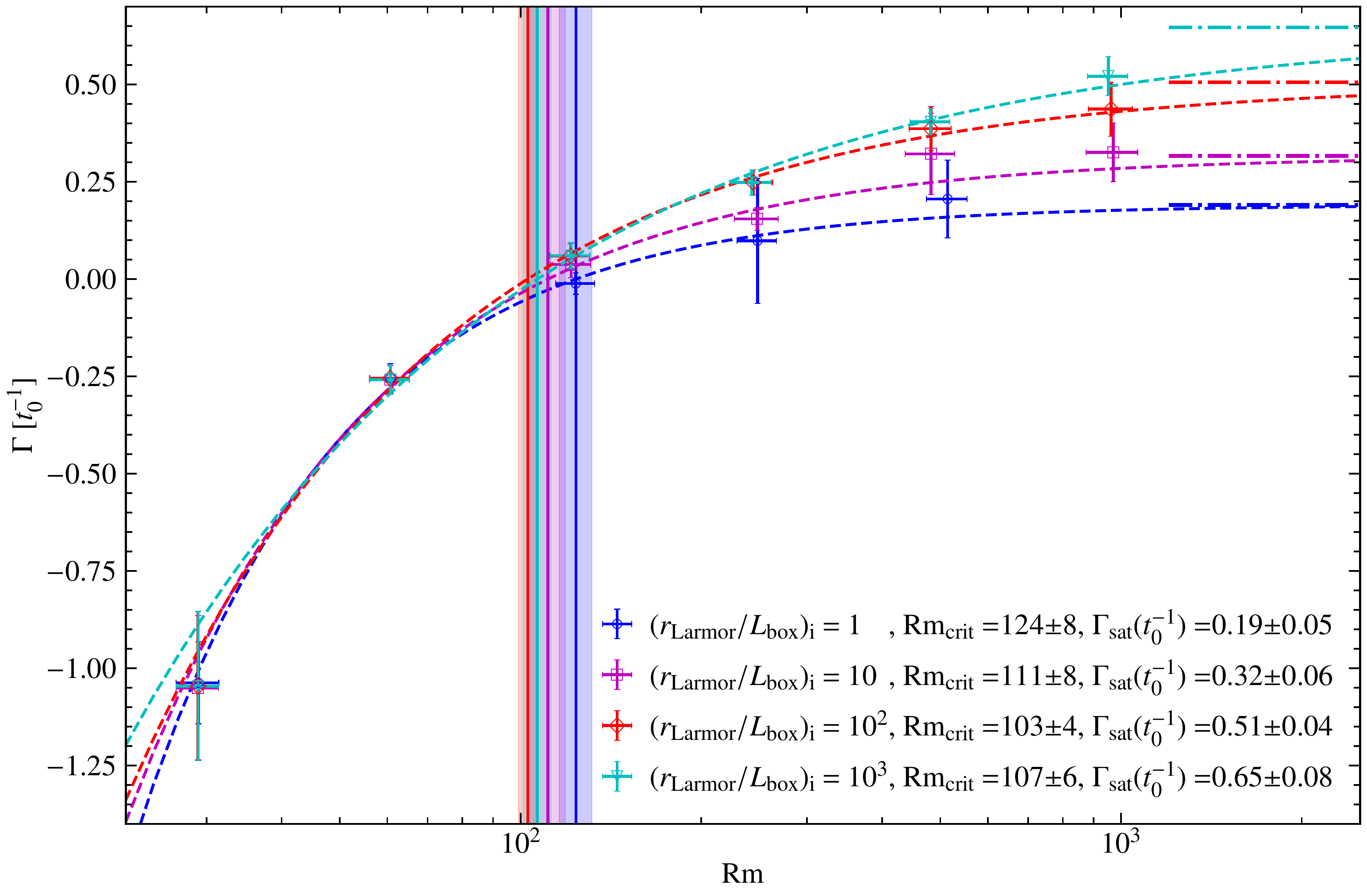}
    \caption{Growth rate ($\Gamma$) of the magnetic energy as a function of the magnetic Reynolds number ($\Rm$) for our collisionless turbulent dynamo models with different initial Larmor ratio ($\initmagnetisation$). The dashed curves are fitted to the simulation data, using \Eq{eqn:Rmfit}. The solid lines show the values of the critical magnetic Reynolds number ($\Rmcrit$, also reported in the legend), and the shaded regions show the error in the measured value of $\Rmcrit$, determined by the fits. The dot-dashed lines on the right-hand corner of the plot depict the saturation level of the growth rate of magnetic energy in the limit of high $\Rm$ ($\Gammasat$) (see \Tab{table:fit_par}).}
    \label{fig:Rmtests_Growthrate}
\end{figure*}

For each value of the initial Larmor ratio, we fit the growth rate as a function of $\Rm$ using the model
\begin{equation}
    \Gamma (\Rm) = \Gamma_{\rm sat} \left[ 1 - \left( \frac{\Rm}{\Rm_{\rm crit}} \right)^{\expcoeff} \right] ,
    \label{eqn:Rmfit}
\end{equation}
where $\Rmcrit$, $\Gamma_{\rm sat}$ and $\alpha$ are fit parameters. $\Gamma_{\rm sat}$ is the saturation level of the growth rate in the limit of $\Rm\to\infty$, motivated by the fast dynamo argument \citep[which suggests that the growth rate of the magnetic field amplification becomes independent of $\Rm$ at very high $\Rm$, see][]{ChildressG1995}, and $\alpha$ is a power-law coefficient. $\Rmcrit$ is the critical magnetic Reynolds number below which magnetic diffusivity dominates, leading to the decay of magnetic fields. When $\Rm > \Rmcrit$, $\Gamma > 0$ and growth of magnetic energy by the collisionless turbulent dynamo is possible. \Figure{fig:Rmtests_Growthrate} shows the growth rate ($\Gamma$) as a function of $\Rm$ for simulations with different $\initmagnetisation$. The solid lines show the values of $\Rmcrit$ and the shaded regions show the error in the measured value of $\Rmcrit$ for different simulation models obtained from fitting using \Eq{eqn:Rmfit}. We also summarise the fit values of $\Rmcrit$, $\Gammasat$, and $\alpha$ for different initial Larmor ratio models in \Tab{table:fit_par}.

\begin{table}
\centering
\caption{Critical magnetic Reynolds number ($\Rmcrit$), saturation level of the growth rate of magnetic energy by the collisionless turbulent dynamo in the limit of high $\Rm$ ($\Gammasat$) and the power law coefficient ($\alpha$) for simulation models with different initial Larmor ratio levels ($\initmagnetisation$) calculated from \Eq{eqn:Rmfit}. }
\begin{tabular}{lccc}
\hline
$\initmagnetisation$ & $\Rmcrit$ & $\Gammasat$ & $\expcoeff$\\
\hline
 $10^{3}$ & 107$\pm$6 & 0.65$\pm$0.08 & $-0.66 \pm 0.12$ \\
 $10^{2}$ & 103$\pm$4 & 0.51$\pm$0.04 & $-0.84 \pm 0.09$ \\
 $10$ & 111$\pm$8 & 0.32$\pm$0.06 & $-1.04 \pm 0.19$ \\
 $1$ & 124$\pm$8 & 0.19$\pm$0.05 & $-1.27 \pm 0.18$ \\
 \hline
\end{tabular}
\label{table:fit_par}
\end{table}

For initially `un-magnetised' plasma ($\initmagnetisation > 1$), we find that the critical value for $\Rm$ is similar for different values of $\initmagnetisation$ we have investigated. For an initially `magnetised' plasma ($\initmagnetisation \lesssim 1$), we find that the critical $\Rm$ for collisionless turbulent dynamo action is marginally higher compared to an initially `un-magnetised' plasma ($\initmagnetisation > 1$). We also find that $\Gammasat$ increases significantly with $\initmagnetisation$. While we control the magnetic Reynolds number of the plasma in our numerical experiments using Ohmic resistivity, the kinetic Reynolds number ($\Rk$) of the collisionless plasma evolves self consistently, i.e., it is determined by the effective viscosity of the plasma, set by interactions between particles and the magnetic field \citep{Kunz+2014,St-Ongeetal2020}. 

In a `magnetised' plasma, we expect the effective viscosity of the plasma to decrease due to the scattering of particles from kinetic instabilities (see \Sec{section:kinetic_instabilities}), which would lead to an increase in the effective kinetic Reynolds number. At a fixed magnetic Reynolds number, for a `magnetised' plasma, the kinetic Reynolds number is thought to be higher when compared to `un-magnetised' plasma. This means the magnetic Prandtl number is smaller for `magnetised' plasma, which can lead to a higher $\Rmcrit$. This effect has been studied for MHD turbulent dynamos \citep{Haugen+2004a,Seta+2020}. Lower magnetic Prandtl numbers for `magnetised' plasma can lead to a decrease in the growth rate of the collisionless turbulent dynamo \citep{Schoberetal2012, Federrathetal2014ApJ}. We further discuss the effective viscosity of collisionless plasma and the effect of magnetic Prandtl number on the growth rate and critical magnetic Reynolds number of the collisionless turbulent dynamo in  \Sec{section:kinetic_instabilities} and \Sec{section:MHDdynamo}.

\section{Initial plasma beta dependence of the dynamo growth rate}
\label{section:initseedstudy}

In this section, we investigate how the initial magnetic-field strength affects the properties of the collisionless turbulent dynamo. For this study, we fix the initial Larmor ratio of the plasma, $\initmagnetisation = 10^{2}$ and the magnetic Reynolds number, $\Rm = 480$, but vary the initial magnetic-to-kinetic energy ratio, $\Einit = 10^{-6}, 10^{-8}$ and $10^{-10}$.

\begin{figure}
    \includegraphics[scale = 0.225]{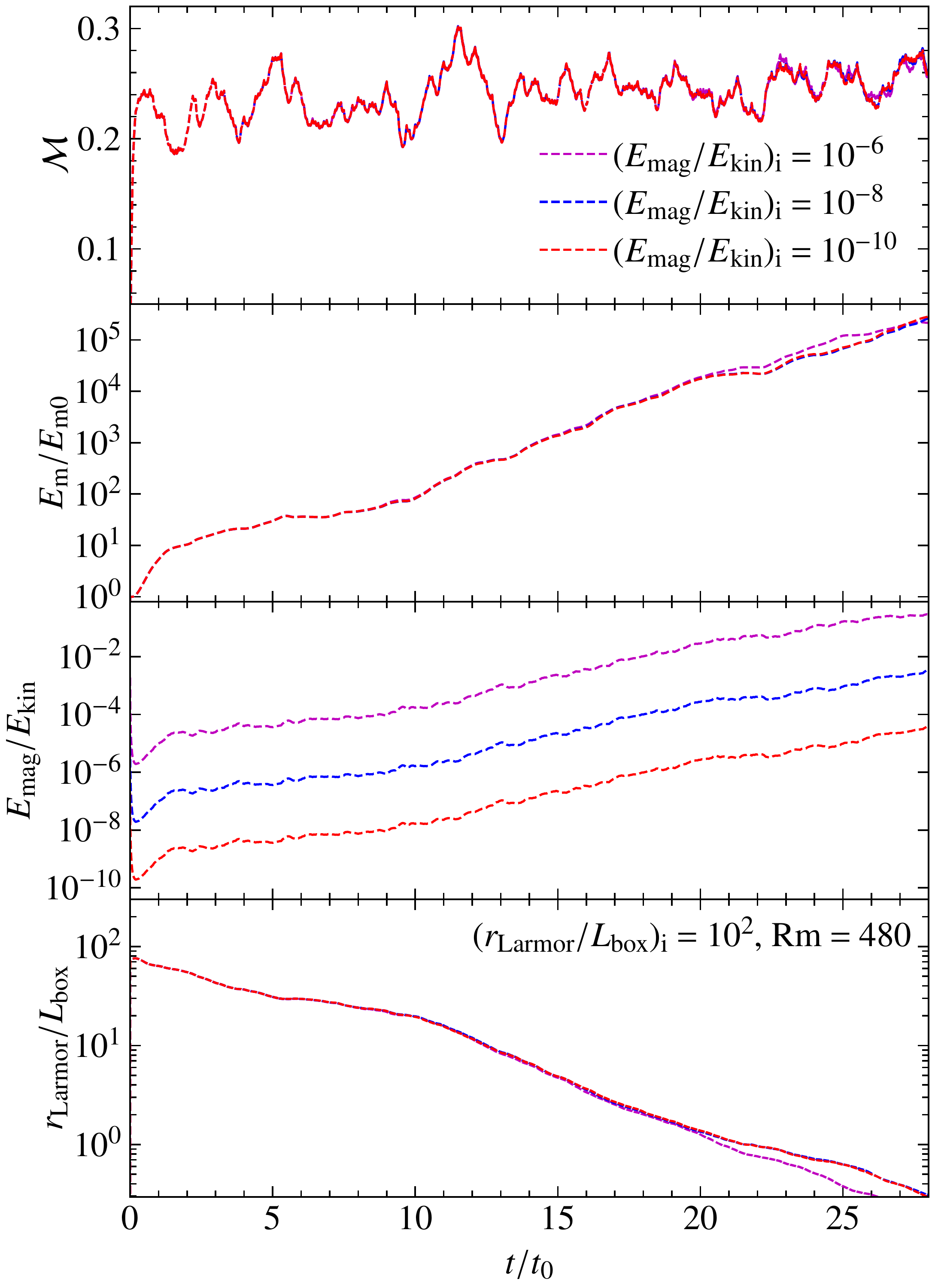}
    \caption{Same as \Fig{fig:Rmstudy_4panelplot_rL1000}, but for an initial ratio of magnetic energy to kinetic energy, $\Einit = 10^{-10}, 10^{-8}$ and $10^{-6}$.  For all the above simulations, we fix the initial Larmor ratio, $\initmagnetisation = 10^{2}$, and the magnetic Reynolds number, $\Rm = 480$ (see \Tab{table:sims}). We see that varying the initial conditions of the plasma does not significantly change the growth rate of the collisionless turbulent dynamo.}
    \label{fig:initratio_4panelplot}
\end{figure}

\Figure{fig:initratio_4panelplot} shows the same as \Fig{fig:Rmstudy_4panelplot_rL100} for collisionless turbulent dynamo simulations with different initial magnetic to kinetic energy ratio ($\Einit$), which describes the initial plasma beta ($\beta_{\rm i}$) as
\begin{equation}
    \beta_{\rm i} = \frac{1}{\Einit \Mach^{2}},
\end{equation}
where the Mach number ($\Mach$) is fixed for all the above simulations. The evolution of the Mach number and the magnetic energy is similar for dynamo simulations with different $\Einit$, as shown by the first and second panels of \Fig{fig:initratio_4panelplot}. We report the measured values of the Mach number and the growth rate in \Tab{table:sims}. We ensure that the average Larmor radius of the charged particles is well resolved for all our simulations, as can be seen from the fourth panel of \Fig{fig:initratio_4panelplot}. 

From the above tests, we conclude that the growth rate of the collisionless turbulent dynamo does not depend on the initial plasma beta at a fixed initial Larmor ratio. This is similar to the behaviour of the MHD turbulent dynamo \citep{Seta&Federrath2020}; we discuss this further in \Sec{section:MHDdynamo}.

\section{Kinetic instabilities in collisionless plasma }
\label{section:kinetic_instabilities}

In collisionless plasma, the thermal pressure can be anisotropic with respect to the local magnetic-field direction \citep{Chew+1956}. This can lead to transport coefficients like viscosity being anisotropic \citep{Braginskii1965}, unlike in MHD for which the pressure is always isotropic. The parallel thermal pressure of the plasma is defined as the projection of the pressure tensor ($\mathsf{P}$) onto the magnetic field, $p_{\parallel} = \mathsf{P}\,\mathbf{:}\,\hat{\mathbf{b}}\hat{\mathbf{b}}$, where $\hat{\mathbf{b}}=\mathbf{B}/B$ is the unit vector in the direction of the local magnetic field. The trace of the pressure tensor can be written as $\text{Tr}(\mathsf{P}) =  p_{\parallel}/3 + 2p_{\perp}/3$, where $p_{\perp}$ is the thermal pressure perpendicular to the magnetic field. As the trace of a tensor is basis invariant, we can obtain the thermal pressure perpendicular to the magnetic field as $p_{\perp} = 3/2(\text{Tr}(\mathsf{P}) - p_{\parallel}/3)$. We further define the pressure anisotropy of the plasma as $\Delta = p_{\perp}/p_{\parallel} - 1$ and the parallel plasma beta as $\beta_{\parallel} = p_{\parallel}/p_{\rm mag}$. For a collisional system, the parallel and perpendicular pressure are made isotropic by collisions, therefore $\Delta = 0$. 

Approximate adiabatic invariance in collisionless plasma couples the thermal motions of the charged particles to changes in the magnetic-field strength. As a result, the pressure tensor becomes anisotropic during the dynamo. In regions with excess perpendicular or parallel thermal pressure, kinetic instabilities can be triggered, causing sharp deflections in the orientation of the local magnetic field. The \emph{mirror instability} destabilises magnetic mirrors in regions where $p_{\perp}/p_{\parallel} - 1 \gtrsim 1/\beta_{\parallel}$. The \emph{firehose instability} can be triggered in the other limit where $p_{\perp}/p_{\parallel} - 1 \lesssim -2/\beta_{\parallel}$, when parallel thermal pressure dominates \citep{Kulsrud2005, Kunz+2014}.

\begin{figure*}
    \centering
    \begin{subfigure}{0.08\textwidth}
        \centering
        \includegraphics[scale=0.43,trim={0.24cm 0.1cm 0.3cm 0.1cm},clip]{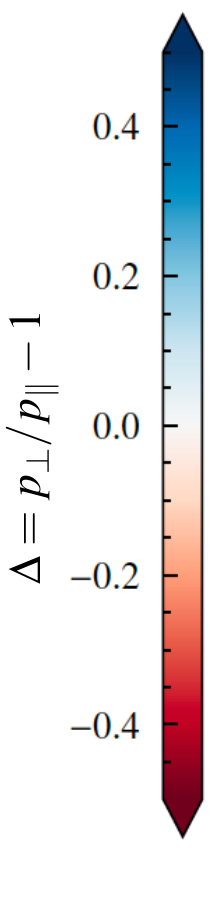}
    \end{subfigure}
    \begin{subfigure}{0.455\textwidth}
        \centering
        \includegraphics[scale=0.299]{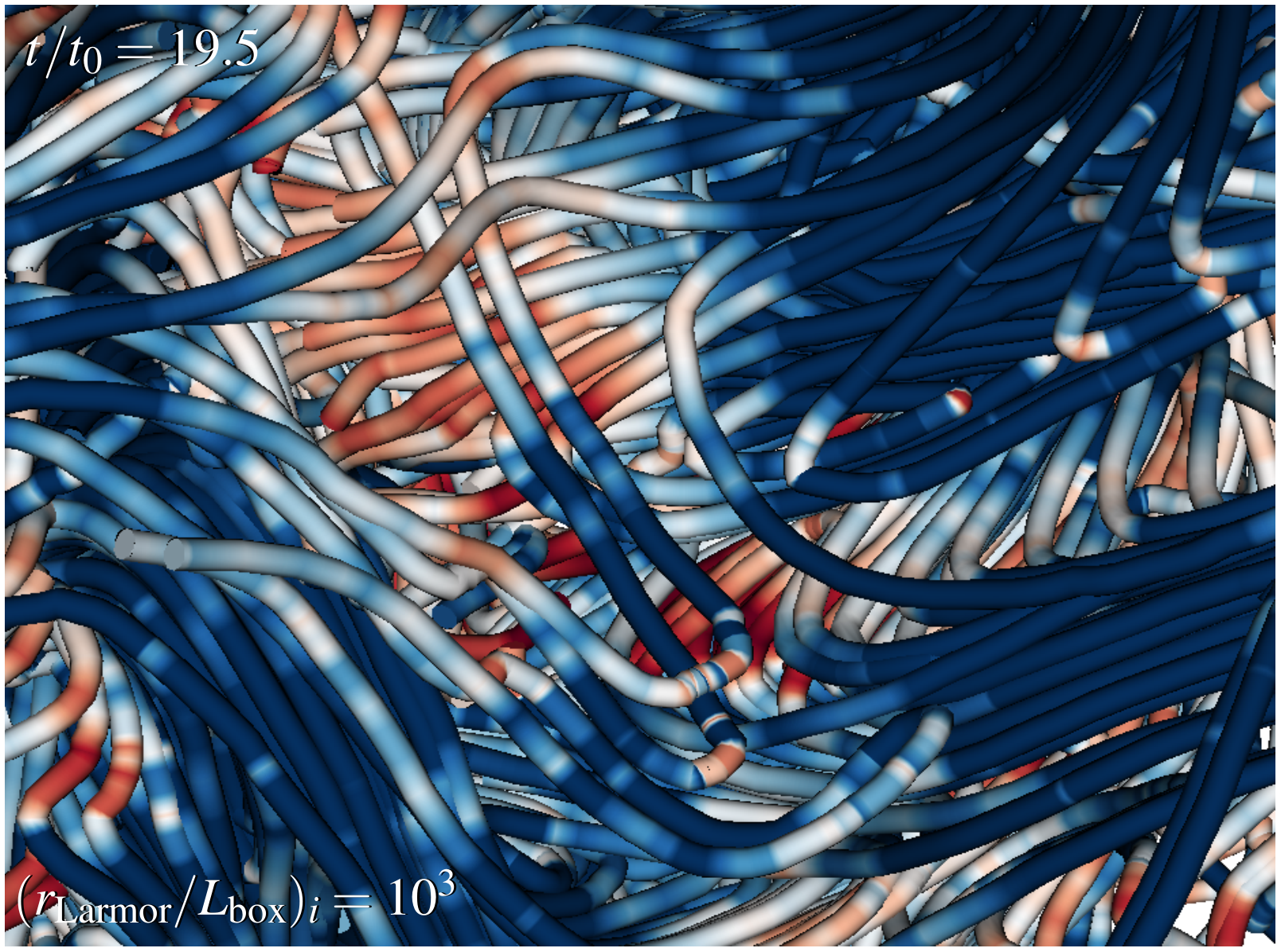}
        \label{fig:rL_1000}
    \end{subfigure}
    \begin{subfigure}{0.455\textwidth}
        \centering
        \includegraphics[scale=0.299]{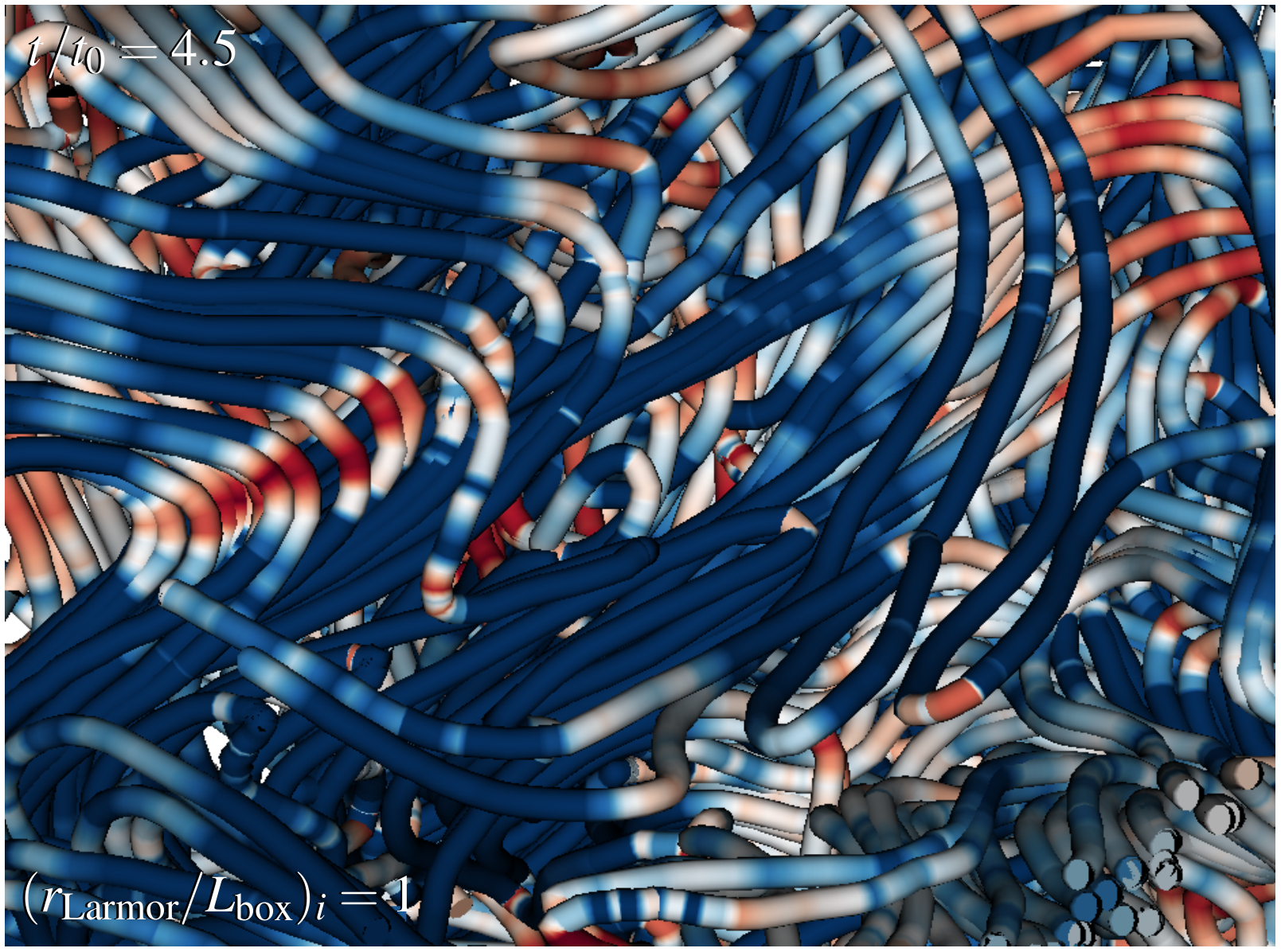}
        \label{fig:rL_1}
    \end{subfigure}
    \caption{3-dimensional rendering of magnetic field streamlines coloured according to the pressure anisotropy ($\sim 0.45 \, L_{\rm box}$). 
    The left panel shows the collisionless turbulent dynamo simulation in the kinematic regime (at $19.5 t_{0}$) with $\initmagnetisation = 10^{3}$, $\Einit = 10^{-8}$, and $\Rm = 482 \pm 37$, while the right panel depicts a simulation in the same regime (at $4.5 t_{0}$) with $\initmagnetisation = 1$, $\Einit = 10^{-8}$, and $\Rm = 514 \pm 40$. Regions coloured in blue have higher perpendicular pressure ($p_{\perp} > p_{\parallel}$) and are possible sites for the mirror instability to act. Regions coloured in red have higher parallel pressure ($p_{\parallel} > p_{\perp}$) and are sites where the firehose instability can be triggered. }
    \label{fig:streamlines}
\end{figure*}

\Figure{fig:streamlines} shows three-dimensional representations of magnetic field streamlines, for simulations with different initial Larmor ratios, $\initmagnetisation = 10^{3}$ and $\initmagnetisation = 1$, where the colour bar corresponds to the pressure anisotropy. These snapshots are shown in the kinematic regime of the collisionless turbulent dynamo and these simulations have $\Einit = 10^{-8}$ and $\Rm \sim 480$. Regions coloured in blue indicate higher perpendicular pressure ($p_{\perp} > p_{\parallel}$), suggesting potential locations for the occurrence of the mirror instability. Regions coloured in red represent higher parallel pressure ($p_{\parallel} > p_{\perp}$), indicating areas where the firehose instability can potentially be triggered. These kinetic instabilities distort the magnetic field on ion-Larmor scales, thereby scattering particles and partially isotropizing the pressure tensor. Hence, kinetic instabilities can regulate the pressure anisotropy by decreasing the viscous stress of collisionless plasma, thereby supplying an effective kinetic Reynolds number. 

\cite{St-Onge&Kunz2018} have studied the distribution of the pressure anisotropy in the initial, kinematic, and saturation phase of the collisionless turbulent dynamo to understand how the mirror and firehose instabilities regulate the dynamo action in the `magnetised' regime. \cite{Rinconetal2016} also find regions of the plasma where the pressure anisotropy satisfies $p_{\perp}/p_{\parallel} - 1 \gtrsim 1/\beta_{\parallel}$ and $p_{\perp}/p_{\parallel} - 1 \lesssim -2/\beta_{\parallel}$, and these two kinetic instabilities can act. In our study, we focus on collisionless turbulent dynamo simulations with higher initial plasma beta to study the dynamo in the kinematic growth phase for a longer period in the `un-magnetised' and `magnetised' regimes.  

\begin{figure*}
    \centering
    \includegraphics[scale = 0.75
    ]{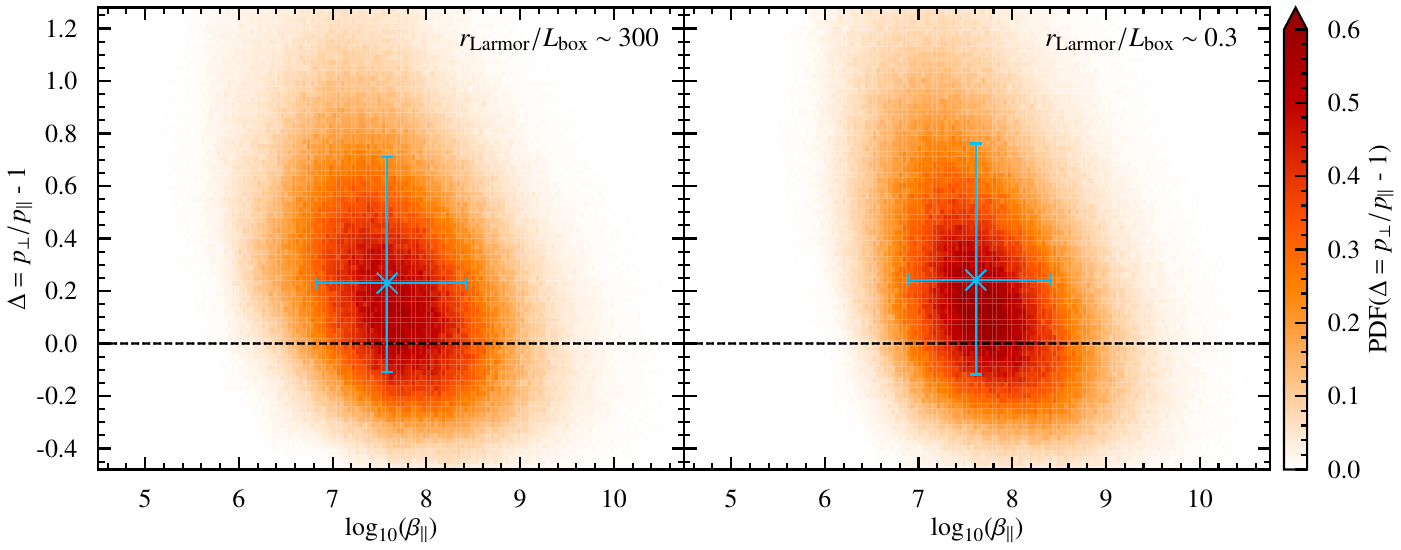}
    \caption{Distribution of the pressure anisotropy ($\Delta = p_{\perp}/p_{\parallel} - 1$) as a function of the parallel plasma beta ($\beta_{\parallel} = p_{\parallel}/p_{\rm mag}$), for simulations with different initial Larmor ratio $\initmagnetisation = 10^{3}$ (left) and $\initmagnetisation = 1$ (right) in the kinematic regime of the collisionless dynamo ($t = 5 \, t_{0}$). The above simulations have $\Rm \sim 480$ and $\Einit = 10^{-8}$. The blue point is the median of the data with the error bars denoting the $16^{\rm th}$ to $84^{\rm th}$ percentile of the data in ${\rm log(\beta_{\parallel})}$ and $p_{\perp}/p_{\parallel} - 1$. The black dotted and dashed curves show the thresholds for the mirror and firehose instability, respectively. For all the above simulations, the median of the pressure anisotropy is greater than zero in the kinematic regime. This indicates that on average in the simulation box, the plasma pressure (or temperature) is anisotropic.}
    \label{fig:anisotropy_2dplot}
\end{figure*}

In \Fig{fig:anisotropy_2dplot}, we plot the distribution of the pressure anisotropy as a function of the parallel plasma beta for simulations with different initial Larmor ratios. We present the data in the kinematic regime (at $5 t_{0}$) for simulations with $\Rm \sim 480$ and $\Einit = 10^{-8}$. The median of the data is represented by the blue point, and the error bars indicate the $16^{\text{th}}$ to $84^{\text{th}}$ percentile in ${\rm log(\beta_{\parallel})}$ and $\Delta$. Additionally, the black dotted and dashed curves illustrate the thresholds for the mirror and firehose instability, respectively. We note that these thresholds are $1/ \beta_{\parallel} \sim 0$ for the high-plasma-beta regime explored in these simulations. The mirror and firehose instabilities enable the dynamo action by scattering the collisionless plasma, thereby increasing the effective collisionality of the plasma. In the kinematic regime, the median of the pressure anisotropy is positive across all simulations with different initial Larmor ratios. We also plot the time evolution of the median value of pressure anisotropy for simulations with different initial Larmor ratios in \Fig{fig:anisotropy_median}.

\begin{table}
\centering
\caption{Time-averaged value of the median of the pressure anisotropy, $\Delta = p_{\perp}/p_{\parallel} - 1$, with the $84^{\text{th}}$ and  $16^{\text{th}}$ percentile values, $(\Delta_{\rm median})^{84^{\text{th}} - 50^{\text{th}} }_{50^{\text{th}} - 16^{\text{th}}}$, and a characteristic magnetic field reversal scale $k_{\vec{B} \times \vec{J}}$ measured in the kinematic regime of the collisionless turbulent dynamo for simulations with different initial Larmor ratio with $\Rm \sim 480$ and $\Einit = 10^{-8}$.}
\begin{tabular}{lcc}
\hline
$\initmagnetisation$ & $\Delta_{\rm median}$ & $k_{\mathbf{B} \times \mathbf{J}} L_{\rm box}/2\pi$
\\
\hline \\
\vspace{0.14 cm}
$10^{3}$ & $0.19_{ -0.32}^{ +0.44}$ & $5.21 \pm 0.28$\\
\vspace{0.14 cm}
$10^{2}$ & $0.19_{ -0.32}^{ +0.44}$ & $5.36 \pm 0.29$\\
\vspace{0.14 cm}
$10$ & $0.18_{ -0.32}^{ +0.44}$ & $5.52 \pm 0.36$\\
\vspace{0.14 cm} 
$1$ & $0.20_{ -0.33}^{ +0.46}$ & $5.88 \pm 0.33$\\
\hline
\end{tabular}
\label{table:anisotropy_val}
\end{table}

We report the median, $16^{\rm th}$, and $84^{\rm th}$ percentile values of the pressure anisotropy, time-averaged in the kinematic regime of the dynamo, for simulations with varying initial Larmor ratios but with fixed $\Rm \sim 480$ and $\Einit = 10^{-8}$ in \Tab{table:anisotropy_val}. Additionally, we report the time-averaged value of the magnetic field reversal scale for these numerical simulations, calculated as \citep{Schekochihin+2004, Seta&Federrath2021a}
\begin{equation}
    k_{\mathbf{B} \times \mathbf{J} }= \mu_{0} \Bigg(\frac{ \langle |\vec{B} \btimes \vec{J}|^{2} \rangle}{\langle B^{4} \rangle}\Bigg)^{1/2}
\end{equation}
in \Tab{table:anisotropy_val}. We find that $k_{\mathbf{B} \times \mathbf{J} }$ increases marginally as the initial Larmor ratio of the plasma decreases. As this scale is similar for collisionless turbulent dynamo experiments with different Larmor ratios, we expect that the Ohmic resistivity sets the dissipation scale in our simulations. As a result, we obtain similar values for the critical magnetic Reynolds number for dynamo action while varying the initial magnetisation of the plasma.  

\section{Comparison with the MHD dynamo}
\label{section:MHDdynamo}
In this section, we compare the properties of the collisionless turbulent dynamo explored in \Sec{section:Rmstudy} and \Sec{section:initseedstudy} to the MHD turbulent dynamo. The MHD turbulent dynamo is a well-studied mechanism, which can efficiently amplify seed magnetic fields by converting turbulent kinetic energy into magnetic energy \citep{Kazantsev1968, Schekochihin+2004, CFetal11, Seta&Federrath2020}. 

\subsection{Critical $\Rm$ for subsonic MHD turbulent dynamo}
The properties of the MHD turbulent dynamo depend on the magnetic Reynolds number ($\Rm$) and the magnetic Prandtl number ($\Pm$) of the plasma. Magnetic field growth by the turbulent dynamo can happen above a certain value of the magnetic Reynolds number known as the critical magnetic Reynolds number for turbulent dynamo action. In \Sec{section:Rmstudy}, we extended this idea to collisionless plasma and found that a critical Reynolds number exists for the collisionless turbulent dynamo action as well. $\Rmcrit$ for the MHD turbulent dynamo depends on the nature of the turbulence driving, the Mach number, and the magnetic Prandtl number of the plasma \citep{Haugen+2004a, CFetal11, Schoberetal2012, Federrathetal2014ApJ, Seta+2020, AchikanathEtAl2021}. 

Numerical studies have shown that $\Rmcrit$ of the MHD turbulent dynamo decreases as $\Pm$ increases \citep{Haugen+2004a,Seta+2020}, because dynamo action is more easily facilitated when the scales at which the kinetic energy dissipates are larger than the scales at which the magnetic energy dissipates \citep{Boldyrev&Cattaneo2004}. Studies have also shown that the MHD turbulent dynamo is feasible at low magnetic Prandtl numbers ($\Pm < 1$) but for higher values of $\Rmcrit$ \citep{Schekochihin+2005a,Iskakov+2007, Schekochihin+2007a, Brandenburg+2018}. Previous studies find that the properties of the collisionless turbulent dynamo are reminiscent of the MHD dynamo in the large Prandtl number regime \citep{Rinconetal2016,St-Onge&Kunz2018, Zhou+2023}. The $\Re$ of the plasma evolves self-consistently and we do not ascertain the $\Re$ in our numerical simulations, therefore it is difficult to predict the $\Pm$ regime of our simulations. We will estimate the $\Re$ in a dedicated upcoming study, which will allow us to understand the $\Pm$ regime of the collisionless turbulent dynamo and better compare our results with the MHD dynamo.

We find that $\Rm_{\rm crit}$ of the collisionless turbulent dynamo in the `un-magnetised' and `magnetised' regime is close to the critical magnetic Reynolds number of the MHD turbulent dynamo for Kolmogorov turbulence \citep{Schoberetal2012}.  We note that the magnetic Reynolds number of the hot ICM is likely to be much higher \citep[$\gg 1$,][]{Schekochihin&Cowley2006} than the values we estimate for $\Rmcrit$ in this study (see \Tab{table:fit_par}). Therefore, if seed magnetic fields are present, it should be easily possible to excite the turbulent dynamo mechanism in the collisionless ICM plasma.

\subsection{Growth rate of the dynamo}
We find from our numerical simulations that the growth rate of the collisionless turbulent dynamo does not depend on the strength of the initial seed magnetic field at a fixed initial Larmor ratio. This is also consistent with the behaviour of the MHD turbulent dynamo, where the growth rate is independent of the strength and nature of the seed field \citep{Seta&Federrath2020}. We conclude that the mechanism that converts small-scale turbulent kinetic energy into magnetic energy is independent of the initial plasma beta when the initial Larmor ratio is fixed. In the context of the hot ICM, if magnetic fields with small strengths from astrophysical or cosmological origins are present, the turbulent plasma can efficiently amplify these magnetic fields via the turbulent dynamo mechanism.

The magnetic Prandtl number determines the scales where the small-scale action of the turbulent dynamo can take place, and studies have shown that the growth rate of the MHD turbulent dynamo increases with $\Pm$ \citep{Schoberetal2012,Federrathetal2014ApJ}. For our collisionless turbulent dynamo simulations at varying $\Rm$ and fixed $\initmagnetisation$, we expect the effective kinetic Reynolds number of the collisionless plasma to be the same. In this case, the magnetic Prandtl number of the collisionless plasma increases as the value of $\Rm$ increases. For each simulation set at fixed $\initmagnetisation$, we find that the growth rate increases with $\Rm$. This is consistent with what \cite{Federrathetal2014ApJ} find for the MHD turbulent dynamo in the supersonic regime. 
The properties of the MHD turbulent dynamo depend on the Mach number and the nature of the turbulent forcing \citep{CFetal11}. In the supersonic regime, the presence of shocks can destroy vorticity modes required to drive the dynamo and decrease the growth rate and saturation efficiency of the turbulent dynamo \citep{Seta&Federrath2022}.

\cite{Rinconetal2016} show that the growth rate of the magnetic energy from the turbulent dynamo in collisionless plasma depends on the initial plasma beta, contrary to what we find. We note that in this study by \cite{Rinconetal2016}, the initial magnetisation of the plasma is changed along with the initial plasma beta, while here, we fix the initial magnetisation of the plasma in our numerical experiments and then vary the initial ratio of magnetic to kinetic energy. When the magnetisation level is fixed, we do not find the growth rate of the dynamo to depend on the initial plasma beta (\Sec{sec:initcond}). The magnetisation level can affect the growth rate of the dynamo (see Fig.~\ref{fig:Rmtests_Growthrate}) as we have discussed earlier and it is important to consider this as an independent parameter for collisionless turbulent dynamo studies.

\citet{Zhou+2023} use fully kinetic simulations to show that, without the presence of initial magnetic fields in a turbulent `un-magnetised' plasma, the Weibel instability can seed magnetic fields and the turbulent dynamo action can grow these fields up to the saturation stage. In the exponential growth phase of the dynamo, where our results may overlap, the growth rate we measure is comparable to what is reported by \citet{Zhou+2023}. \citet{St-Onge&Kunz2018} report the growth rate of the collisionless dynamo in the `magnetised' regime $\approx 0.15 \Vturb/L$ in the exponential growth phase. This is similar to the growth rate we find in the `magnetised' case, $\sim 0.21 t_{0}^{-1}$ for $\Rm \sim 510$ and $\Einit = 10^{-8}$. \citet{Rinconetal2016} find the growth rate of the dynamo $\approx 0.16 \, \Vturb/L$ in their `un-magnetised' and high-beta simulations which is lower than what we find in the `un-magnetised' regime. We note that there are caveats to this comparison as the $\Rm$ differs greatly across these studies.

\section{Conclusions}
\label{section:conclusions}
We study the properties of the collisionless turbulent dynamo in the kinematic growth phase using hybrid-kinetic particle-in-cell simulations with the FLASH code. We solve the hybrid-kinetic equations with a turbulent driving field modelled by an Ornstein--Uhlenbeck process. We use a novel cooling method to cool the collisionless plasma, in order to maintain a constant temperature and to allow for steady-state turbulence at any target sonic Mach number. We change the magnetic Reynolds number ($\Rm \approx 30 - 960$) of the collisionless plasma for four different values of initial magnetisation ($\initmagnetisation$) in the `magnetised' and `un-magnetised' regime and find that a critical value for the magnetic Reynolds number exists in both regimes and is comparable to that for the ${\rm Pm}\gtrsim 1$ MHD turbulent dynamo. We also find that the growth rate of the collisionless turbulent dynamo increases with the magnetic Reynolds number, irrespective of the initial magnetisation. In the `un-magnetised' regime, we find that the critical value of the magnetic Reynolds number, $\Rmcrit = 107 \pm 6$ for $\initmagnetisation = 10^{3}$, $\Rmcrit = 103 \pm 4$ for $\initmagnetisation = 10^{2}$ and $\Rmcrit = 111 \pm 8$ for $\initmagnetisation = 10$. In the `magnetised' regime with $\initmagnetisation = 1$, we find that $\Rmcrit = 124 \pm 8$. 

We also examine how the strength of the seed magnetic field affects the growth rate of the collisionless turbulent dynamo by varying the initial magnetic energy to kinetic energy ratio, $\Einit$, while fixing the initial Larmor ratio of the plasma $\initmagnetisation = 10^{2}$. We find that the growth rate does not depend on $\Einit$, similar to the MHD turbulent dynamo. 

We study the distribution and evolution of the pressure anisotropy of the collisionless plasma ($\Delta = p_{\perp}/p_{\parallel} - 1$) for different values of initial magnetisation during the kinematic regime of the dynamo. We find that the evolution of the pressure anisotropy is similar for all our simulation models. The median pressure anisotropy, $\Delta$, remains approximately 0.2 throughout the kinematic regime in all the simulations we study. Additionally, we visualise regions where the mirror and firehose instabilities, which increase the effective collisionality of the plasma, can be present during the growth phase of the dynamo. We also compare the critical Reynolds number and growth rate of the collisionless dynamo with those of the MHD turbulent dynamo. We will investigate the effective collisionality to understand the kinetic Reynolds number and behaviour of the collisionless turbulent dynamo in the saturation regime in future studies.
 
\section{Acknowledgements}
R.~A.~C.~acknowledges that this work was supported by an NCI HPC-AI Talent Program 2023 Scholarship, with computational resources provided by NCI Australia (project gp08), an NCRIS-enabled capability supported by the Australian Government.
C.~F.~acknowledges funding provided by the Australian Research Council (Future Fellowship FT180100495 and Discovery Project DP230102280), and the Australia-Germany Joint Research Cooperation Scheme (UA-DAAD). M.~W.~K.~was supported in part by NSF CAREER Award No.~1944972. We further acknowledge high-performance computing resources provided by the Leibniz Rechenzentrum and the Gauss Centre for Supercomputing (grants~pr32lo, pr48pi and GCS Large-scale project~10391), the Australian National Computational Infrastructure (grant~ek9) and the Pawsey Supercomputing Centre (project~pawsey0810) in the framework of the National Computational Merit Allocation Scheme and the ANU Merit Allocation Scheme. This work has greatly benefited from discussions held at the scientific program on "Magnetic Field Evolution in Low Density or Strongly Stratified Plasmas" held at the Nordic Institute for Theoretical Physics in 2022. The simulation software, \texttt{FLASH}, was in part developed by the Flash Centre for Computational Science at the Department of Physics and Astronomy of the University of Rochester.

\section*{Data Availability}
Simulation results would be shared on reasonable request to the corresponding author.



\bibliographystyle{mnras}
\bibliography{Reffiles_radhika.bib} 




\appendix

\section{Simulations with different initial Larmor ratio}
\label{Appendix:Larmor_ratio}

\begin{figure}
    \includegraphics[scale = 0.225]{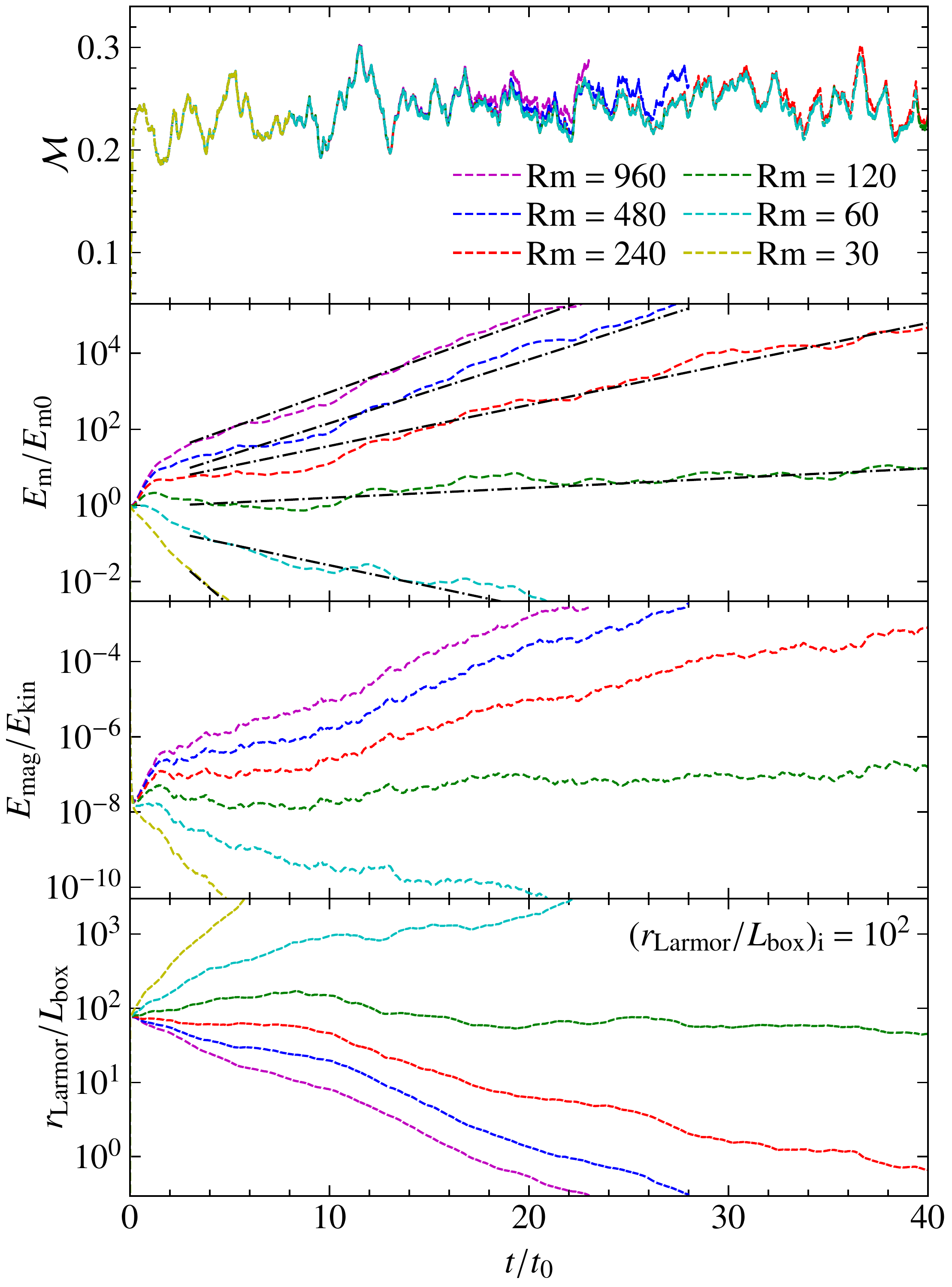}
    \caption{Same as \Fig{fig:Rmstudy_4panelplot_rL1000}, but for an initial Larmor ratio of $\initmagnetisation = 10^{2}$ (see \Tab{table:sims}). }
    \label{fig:Rmstudy_4panelplot_rL100}
\end{figure}

\begin{figure}
    \includegraphics[scale = 0.225]{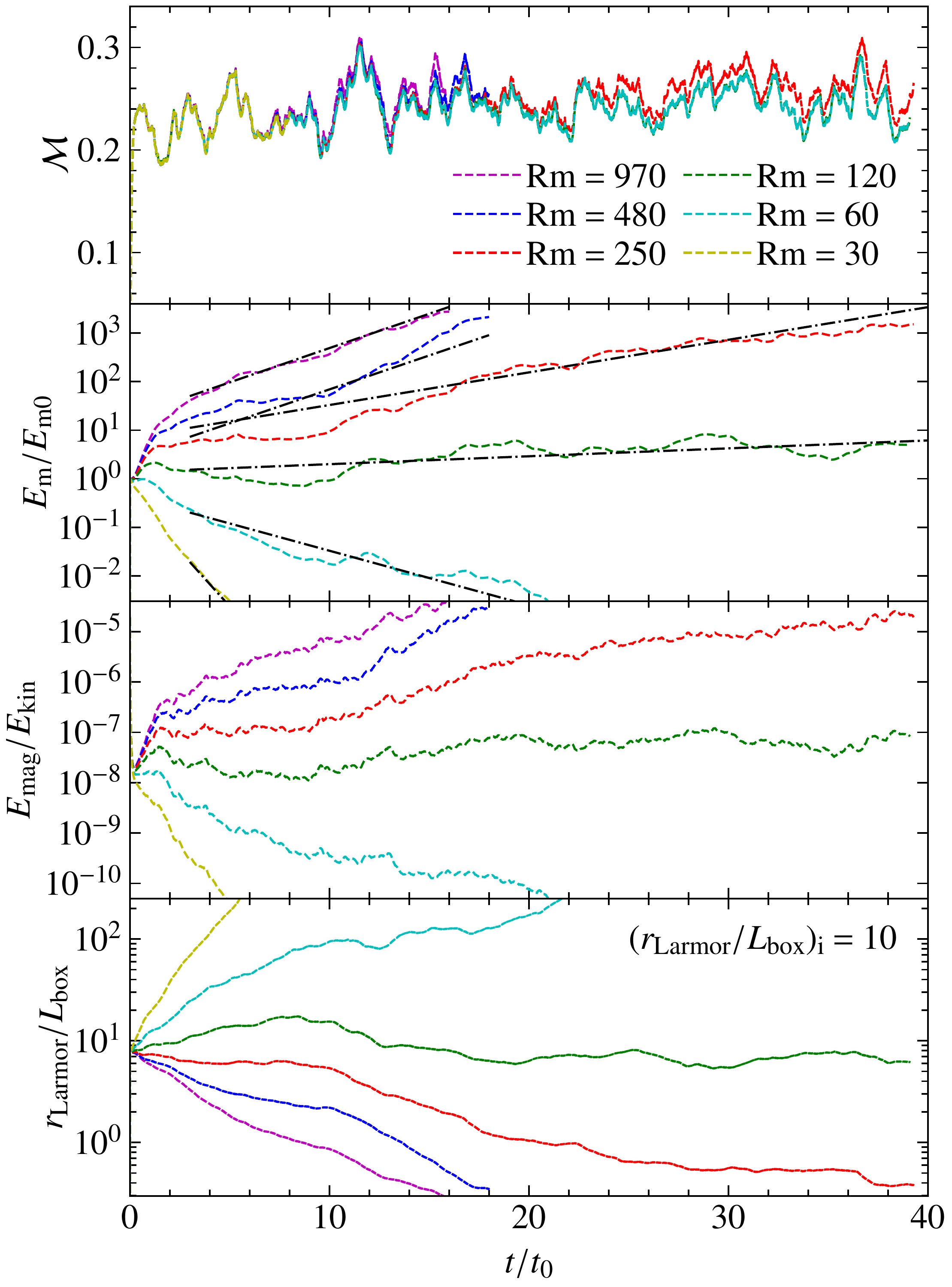}
    \caption{Same as \Fig{fig:Rmstudy_4panelplot_rL1000}, but for an initial Larmor ratio, $\initmagnetisation = 10$ (see \Tab{table:sims}).}
    \label{fig:Rmstudy_4panelplot_rL10}
\end{figure}

\begin{figure}
    \includegraphics[scale = 0.225]{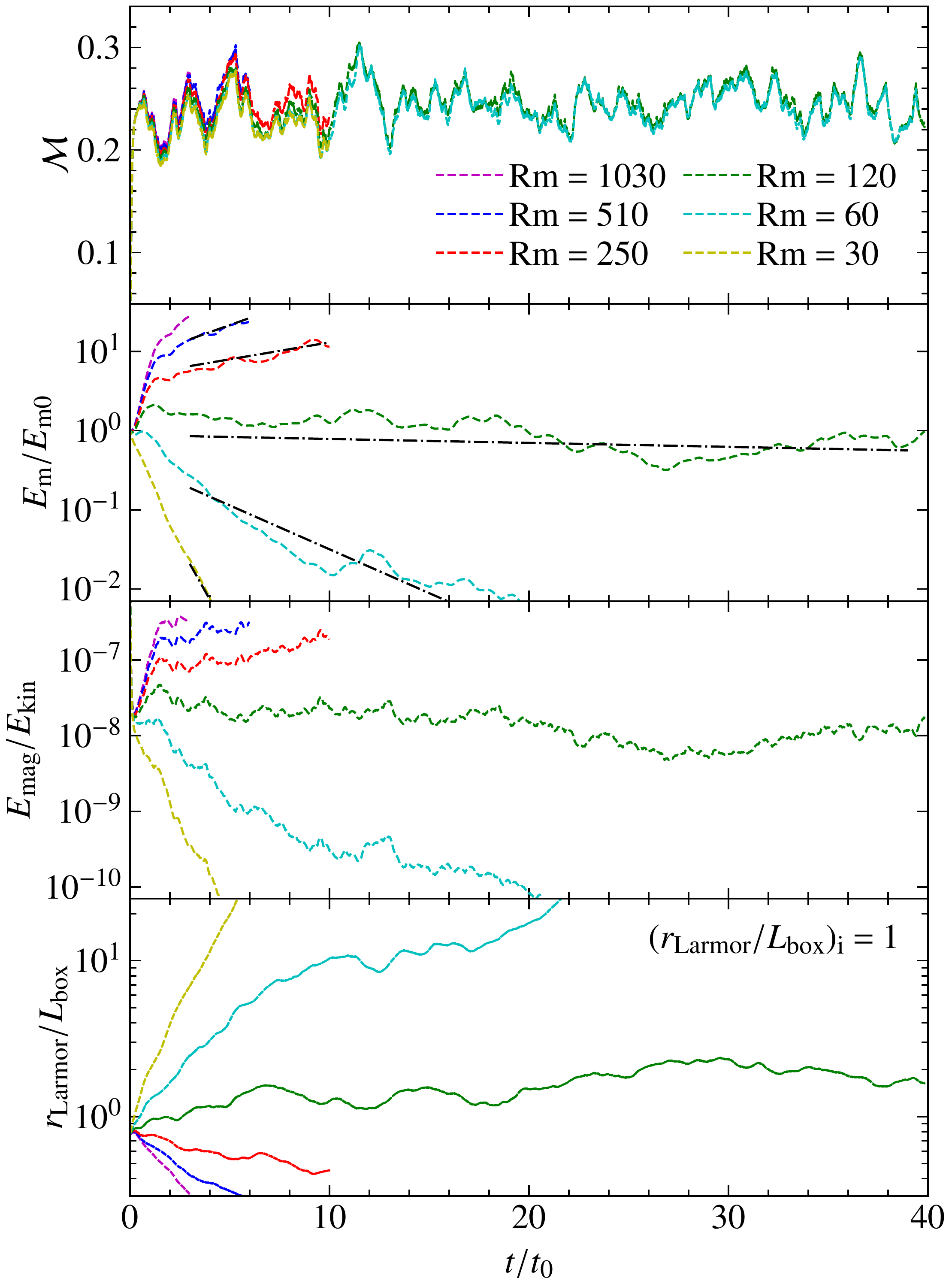}
    \caption{Same as \Fig{fig:Rmstudy_4panelplot_rL1000}, but for an initial Larmor ratio, $\initmagnetisation = 1$ (see \Tab{table:sims}).}
    \label{fig:Rmstudy_4panelplot_rL1}
\end{figure}

We perform the same experiment changing the value of the magnetic Reynolds number with a different initial Larmor ratio, $\initmagnetisation = 10^{2}, 10$ and 1 and present the time evolution of these simulations in \Fig{fig:Rmstudy_4panelplot_rL100}, \ref{fig:Rmstudy_4panelplot_rL10} and \ref{fig:Rmstudy_4panelplot_rL1} respectively. The results are qualitatively similar to what we find for the study with $\initmagnetisation = 10^{3}$. For sufficiently high values of $\Rm$, there is a clear exponential growth phase, while for values of $\Rm$ below a critical value ($\Rmcrit$ determined in \Sec{section:Rmstudy}), the magnetic field decays. We note that, as we decrease the initial Larmor ratio, the range within which the Larmor radius of particles can be resolved also decreases, resulting in a smaller range for the simulation.

\section{Effect of random seed field on the evolution of the dynamo}
\label{Appendix:randomseed}

\begin{figure}
    \includegraphics[scale = 0.225]{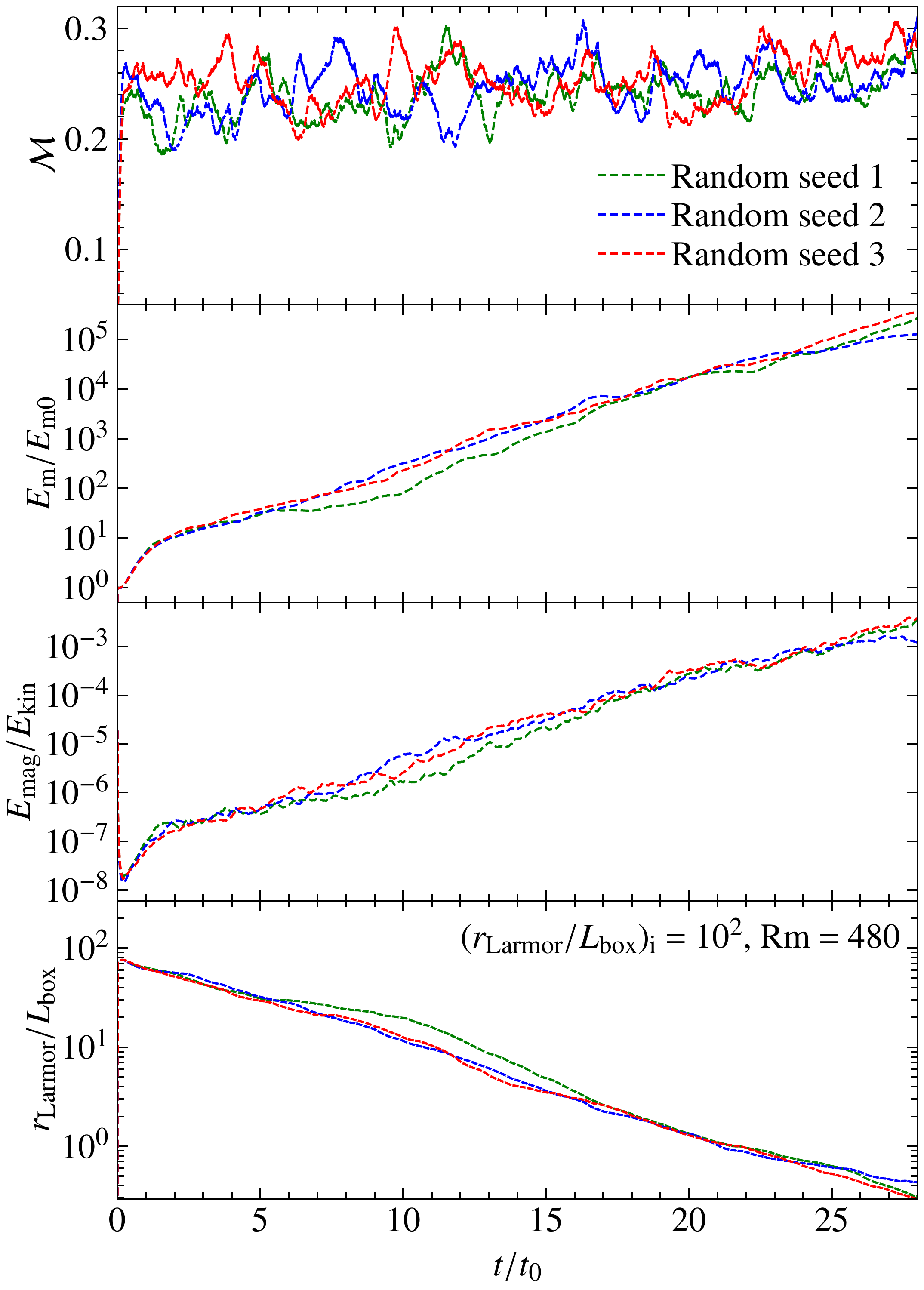}
    \caption{Same as \Fig{fig:Rmstudy_4panelplot_rL1000}, but for different random number seeds. The above simulations have magnetic Reynolds number, $\Rm = 480$ and $\Einit = 10^{-8}$ (see \Tab{table:sims}, Ser. No. 11, 26 and 27). The value of the random turbulent seed has some effect on the evolution of the Mach number and also affects the growth of magnetic energy, which is quantified here. For the growth rate, the variation is $\sim 10 \%$ for models with different random seeds.}
    \label{fig:randomseed_test}
\end{figure}

The turbulent forcing sequence, $\textbf{f}$, is initialised with a random number generator seed. This seed value determines the evolution of the turbulence driving field. The amplification of the magnetic energy is somewhat sensitive to fluctuations in the turbulent speed, $\Vturb$, as a result of this random driving sequence, and this can affect the locally measured growth rates of the dynamo, in particular when only a relatively short time interval is available for averaging (see \App{Appendix:growthrateerr}). To test this, we run the simulation model with $\Rm = 480$, $\Einit = 10^{-8}$ and $\initmagnetisation = 10^{2}$ with three different random seed values, and present the results in \Fig{fig:randomseed_test}.

We find that across different turbulent realisations, the local variations in turbulence, such as the crests and troughs in Mach number, vary to some degree, and this leads to variations in the magnetic field amplification as well. As expected, if averaged over long periods in the kinematic regime of the dynamo, the average Mach number and the measured growth rate of the dynamo are independent of the seed (see \Tab{table:sims}), but these variations can be important for short periods, as can be seen from \Fig{fig:randomseed_test}. We note that it is important to consider the effect of the random seed used in the turbulent forcing field for numerical studies of driven turbulence and turbulent dynamo experiments.

\section{Accounting for systematic error in growth rate measurements}
\label{Appendix:growthrateerr}

We measure the growth rate of the collisionless turbulent dynamo by fitting an exponential curve to the magnetic energy growth in the kinematic regime of the dynamo. However, the interval over which the fit is performed is non-standardised (different for different parameters) and this can lead to a systematic error in the measured value of the growth rate. To understand and mitigate this, we devise a new method that automatically finds the appropriate error of the growth rate based on the available data (length of time evolution). We develop this method by measuring the growth rate of the collisionless dynamo model in one of our models: $\Rm = 240$, $\Einit = 10^{-8}$ and $\initmagnetisation = 10^{3}$, for varying time intervals. However, the results of this are transferable to all of our models and may be more generally applied when measuring turbulent dynamo growth rates. Development of steady-state turbulence takes $\lesssim 3 \, t_{0}$ in our simulations and thus we generally do not include this initial phase in the growth rate measurement.

\begin{figure}
    \includegraphics[scale = 0.196]{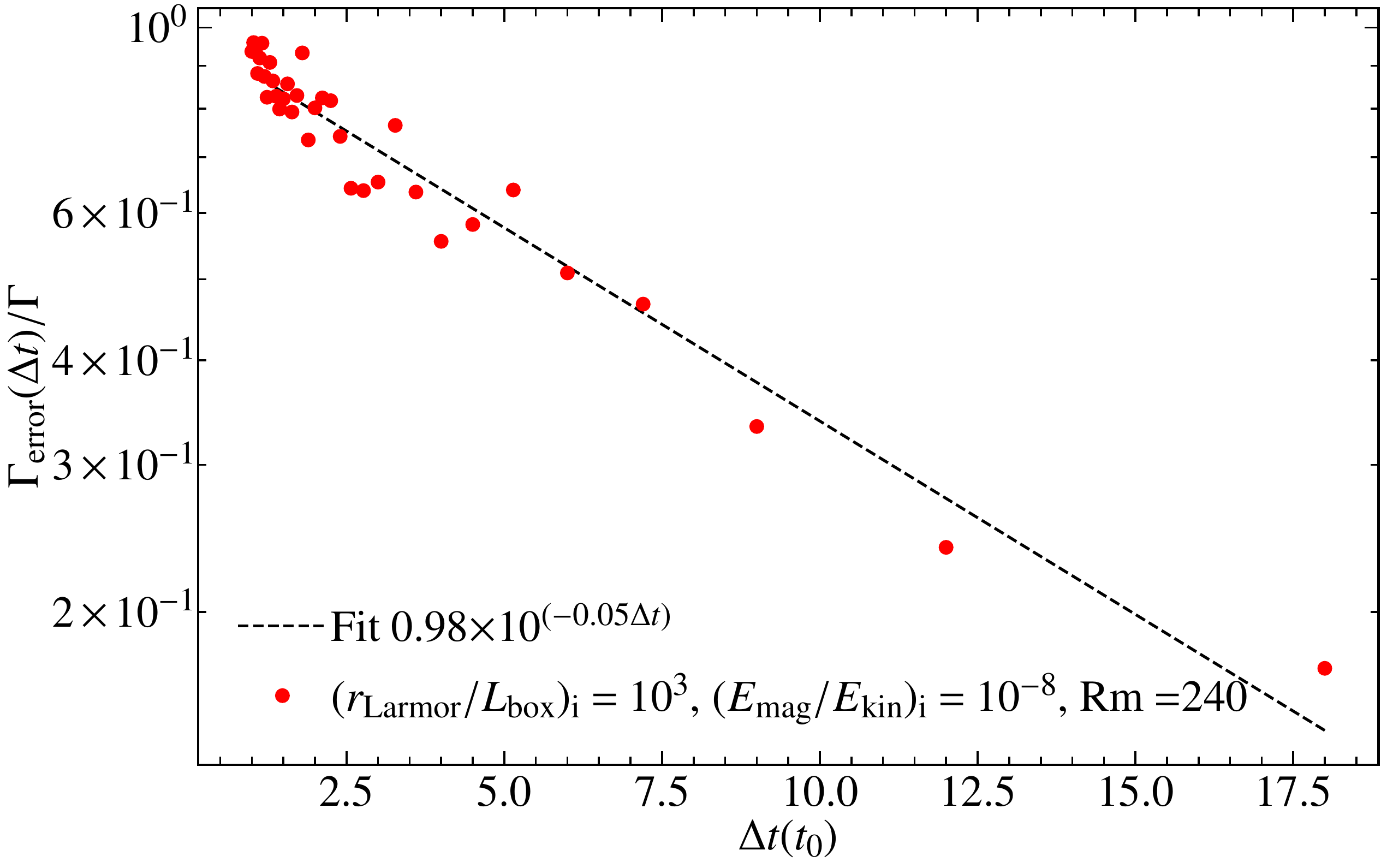}
    \caption{Relative error in the dynamo growth rate as a function of the sub-interval time measured for a representative model from the main part of the study. We see that the error decreases as the time interval for averaging is increased. Using the fitted function, shown in the dotted black line (see legend), we provide a standardised error for the dynamo growth rate, given by \Eq{eq:Gamma_error}).}
    \label{fig:growthrate_error}
\end{figure}

We start by dividing the kinematic regime of this particular simulation (the time interval $3 - 39 \, t_{0}$) of total length $D t = 36 \, t_{0}$ into two sub-intervals of $\Delta t = 18 \, t_{0}$ ($D t/2$) each, three sub-intervals of $\Delta t = 12 \, t_{0}$ each, etc., up to 36 sub-intervals each, with $\Delta t = 1 \, t_{0}$, and fit an exponential curve to each sub-interval.
The growth rate, $\Gamma$, we report for all our simulations is the mean of the sub-intervals with $\Delta t = 1 \, t_{0}$. We compute the standard deviation of the growth rates measured for each set of sub-intervals  ($\Gamma_{\rm error}(\Delta t)$).
\Figure{fig:growthrate_error} shows that the error in the growth rate increases as $\Delta t$ decreases because, for small sub-intervals, the growth rate measurement is strongly influenced by local features of the turbulence, leading to higher variation across sub-intervals and larger errors. Over larger $\Delta t$, the response of the magnetic energy growth to local turbulent features is averaged out, naturally leading to smaller errors.

Using this information, we fit a decaying power law to the relative error in the growth rate ($\Gamma_{\rm error} (\Delta t)/\Gamma$) as a function of the time interval used for averaging ($\Delta t$), and obtain the fitted function $f (\Delta t) \propto 10^{-0.05 \Delta t}$ for the relative error in the dynamo growth rate as a function of $\Delta t$. For each simulation, we measure the error in the growth rate at $\Delta t = 1 \, t_{0}$, i.e., $\Gamma_{\rm error} (\Delta t = 1 \, t_{0})$, and use the following function to calculate the error in the growth rate, such that the final result for the error is independent of $\Delta t$,
\begin{equation} \label{eq:Gamma_error}
    \Gamma_{\rm error} = \frac{f(\Delta t = Dt/2)}{f(\Delta t = 1 \, t_{0})} \Gamma_{\rm error} (\Delta t = 1 \, t_{0}),
\end{equation}
where $f(\Delta t = Dt/2)$ is the value of the function evaluated when the total kinematic regime is divided into two sub-intervals. This method mitigates the systematic error in choosing the time interval over which fitting for the growth rate is done. We use this functional form to estimate errors for all our runs with varying fit intervals.

\section{Effect of resolution on numerical experiments}
\label{Appendix:res_test}

To illustrate the convergence of our collisionless dynamo solutions with both grid and particle resolution, we repeat our numerical experiments with different grid resolutions and number of particles per cell for the simulation model with $\Rm = 480$ and $\Einit = 10^{-8}$.

\begin{figure}
    \includegraphics[scale = 0.225]{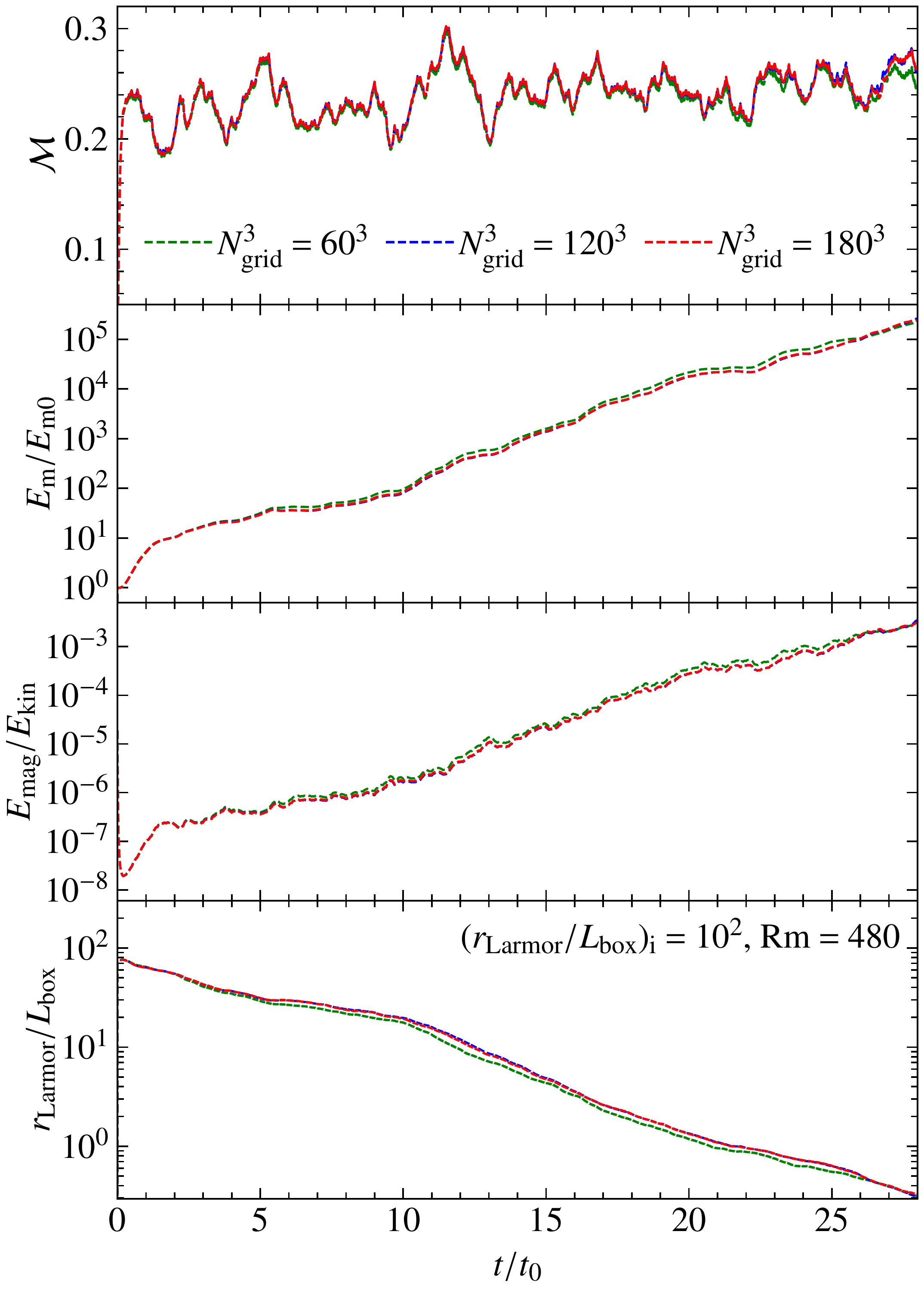}
    \caption{Same as \Fig{fig:Rmstudy_4panelplot_rL100}, but for different grid resolution ($\ngrid^{3}$) of $60^{3}$, $120^{3}$ and $180^{3}$ cells. We find that the evolution of the Mach number and the growth rate of the dynamo is converged at a grid resolution of $120^{3}$ cells, given the Rm (resistivity) chosen in this experiment. Thus, we generally capture the grid resolution requirements for convergence given a target resistivity (see Sec~\ref{sec:resihype}).}
    \label{fig:gridres_test}
\end{figure}

\Figure{fig:gridres_test} shows the time evolution of the dynamo for simulations with a grid resolution of $60^{3}$, $120^{3}$, and $180^{3}$, with a fixed number of particles-per-cell ($\nppc = 100$). We do not find any significant difference in the Mach number and growth rate of the dynamo as we change the numerical grid resolution.

\begin{figure}
    \includegraphics[scale = 0.225]{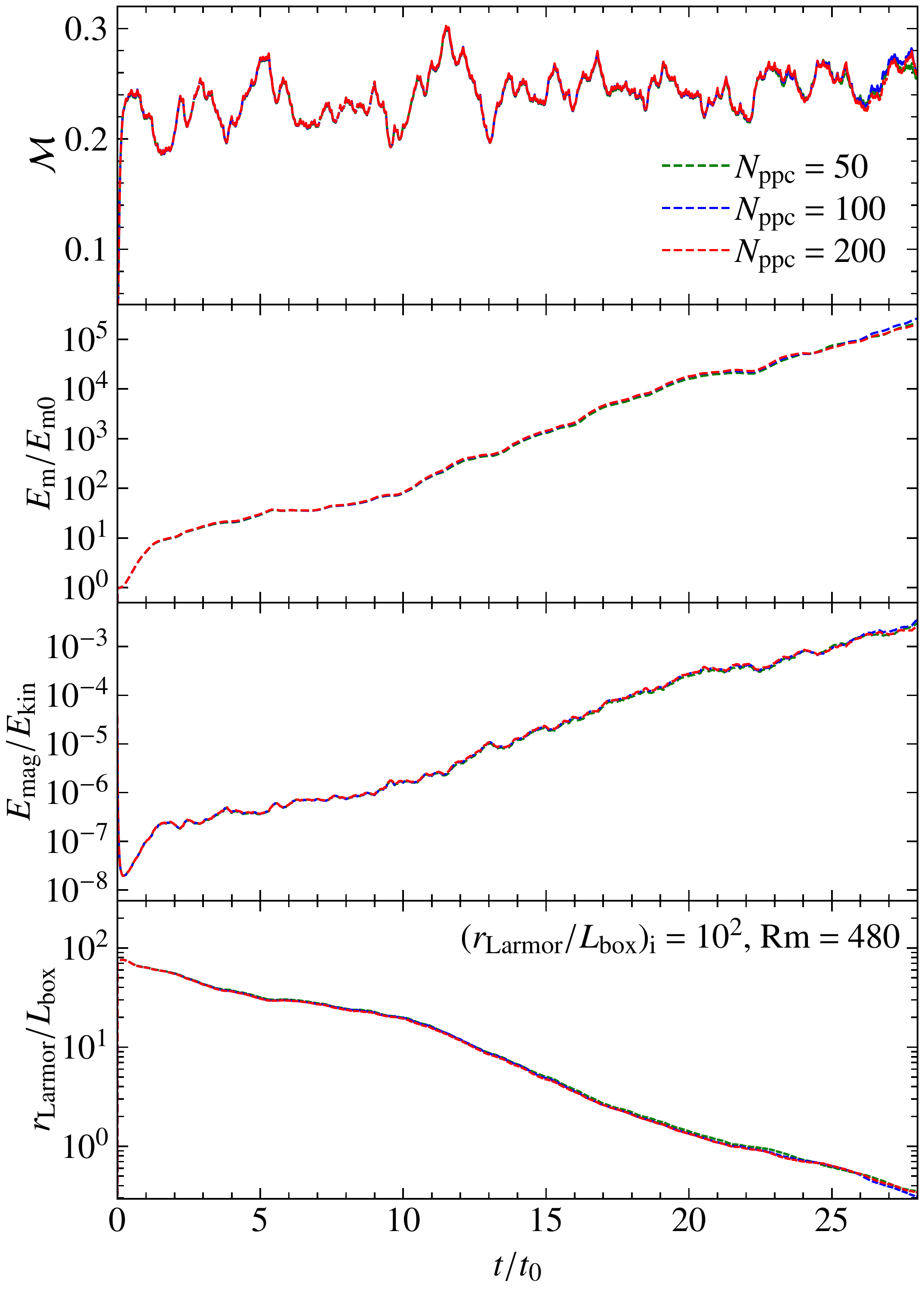}
    \caption{Same as \Fig{fig:Rmstudy_4panelplot_rL100}, but for particles-per-cell counts of $\nppc = 50, 100$ and 200. We do not find any significant dependence on $\nppc$ for the plasma parameters studied here.}
    \label{fig:ppcres_test}
\end{figure}

Next, we fix the grid resolution to $120^{3}$ and vary the number of particles-per-cell to $\nppc = 50, 100$ and 200. The time evolution of these simulations is shown in \Fig{fig:ppcres_test}. We do not find any significant variation in the Mach number and growth rate of the dynamo for different particle resolutions with the current physical parameters of the plasma.

\section{Magnetic power spectra}
\label{Appendix:mags_spectra}

\begin{figure}
    \includegraphics[scale = 0.34]{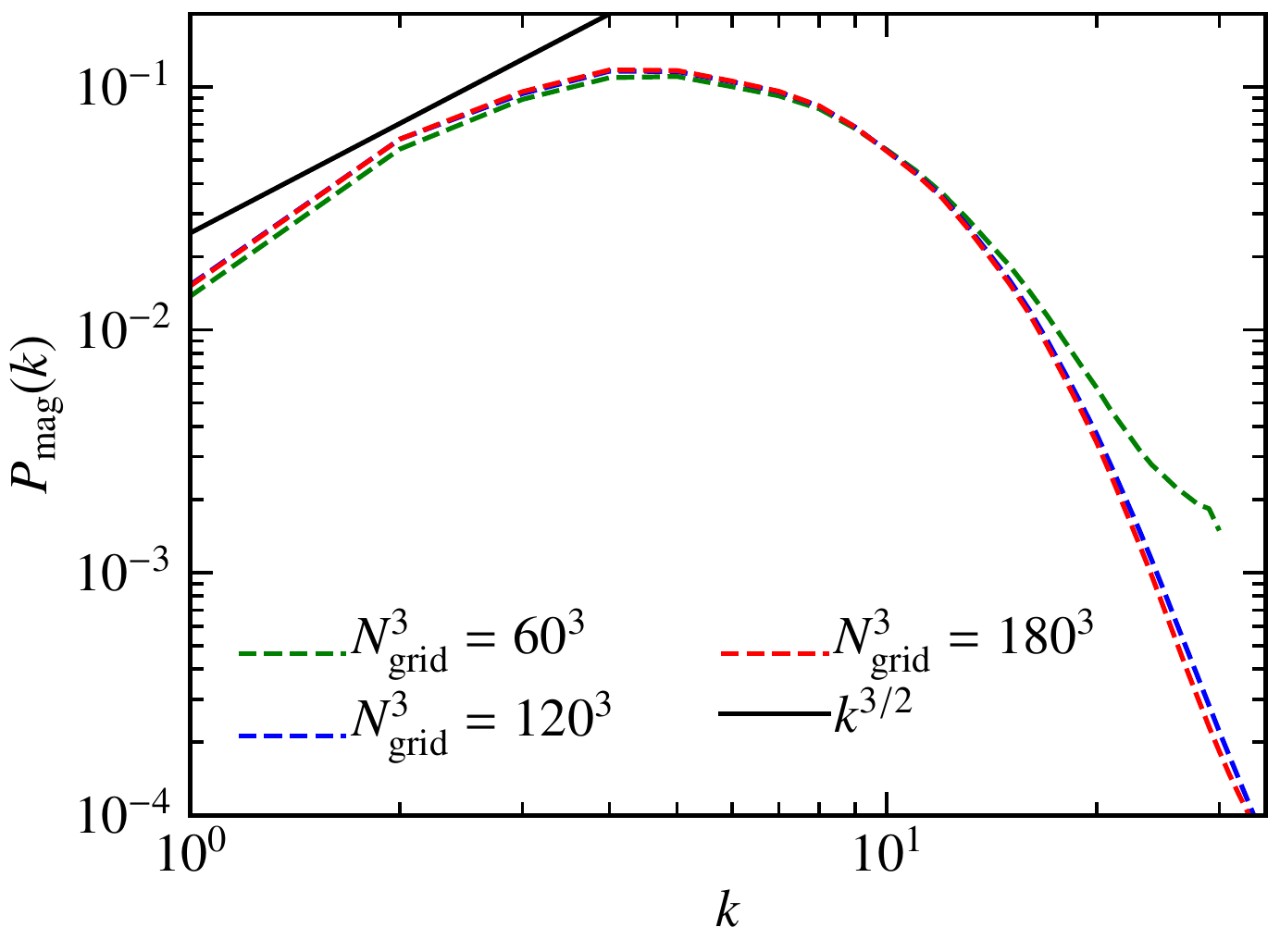}
    \caption{Time-averaged magnetic power spectra of the collisionless turbulent dynamo in the kinematic regime with magnetic Reynolds number, $\Rm = 480$, initial magnetic to kinetic energy ratio, $\Einit = 10^{-8}$, initial Larmor ratio, $\initmagnetisation = 10^{2}$, and $\nppc = 100$ for different grid resolutions $\ngrid^{3} = 60^{3}, 120^{3}$ and $180^{3}$. The $k^{3/2}$ scaling, characteristic of the MHD dynamo \citep{Kazantsev1968}, is shown for comparison.}
    \label{fig:mag_spec}
\end{figure}

We plot the time-averaged magnetic power spectra for our collisionless turbulent dynamo simulations in the exponential growth regime with fixed magnetic Reynolds number, $\Rm = 480$, initial Larmor ratio, $\initmagnetisation = 10^{2}$, initial magnetic to kinetic energy ratio, $\Einit = 10^{-8}$, and $\nppc = 100$  for different grid resolutions, $\ngrid^{3} = 60^{3}, 120^{3}$ and $180^{3}$ in \Fig{fig:mag_spec}. We find that the magnetic power spectra on larger scales converge for different grid resolutions and are visually consistent with the $k^{3/2}$ scaling characteristic of the MHD dynamo for all the grid resolutions \citep{Kazantsev1968}.

\section{Evolution of pressure anisotropy}
\label{Appendix:evolution_presanis}

\begin{figure*}
    \centering
    \includegraphics[scale = 0.7]{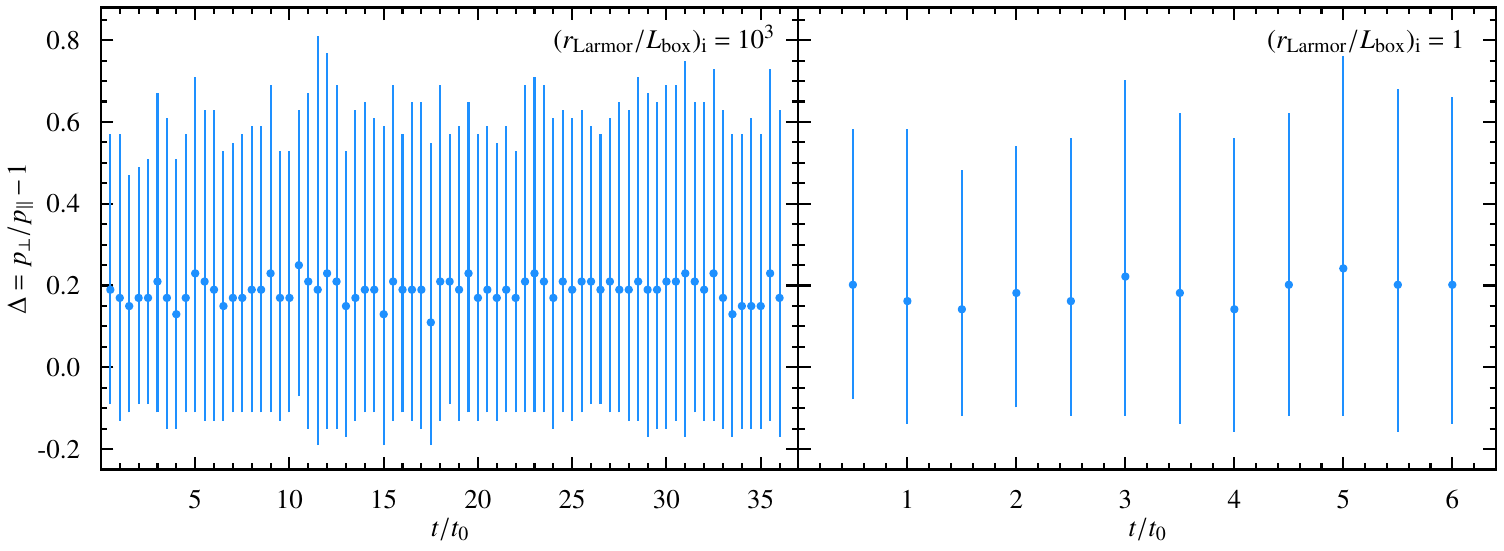}
    \caption{Evolution of the median of the pressure anisotropy ($\Delta = p_{\perp}/p_{\parallel} - 1
    $) as a function of time, normalised to the turbulent turnover time ($t_{0}$), for simulations with different initial Larmor ratio $\initmagnetisation = 10^{3}$ (left) and $\initmagnetisation = 1$ (right).  The above simulations have $\Rm \sim 480$ and $\Einit = 10^{-8}$. The lower and upper errors on each point show the $16^{\rm th}$ to $84^{\rm th}$ percentile range of the pressure anisotropy. In all the above simulations with varying initial Larmor ratios, the median of the pressure anisotropy $\sim 0.2$ during the kinematic regime.}
    \label{fig:anisotropy_median}
\end{figure*}

In \Fig{fig:anisotropy_median}, we plot the time evolution of the median value of pressure anisotropy for simulations with varying initial Larmor ratios. The lower and upper error bars represent the $16^{\text{th}}$ to the $84^{\text{th}}$ percentile values of the pressure anisotropy respectively. The median of the pressure anisotropy is $\Delta \sim 0.2$ throughout the kinematic regime for all the simulations we study and is similar across simulations with different initial magnetisations. We report the time-averaged value of the pressure anisotropy in the kinematic regime of the collisionless turbulent dynamo for simulations with different initial Larmor ratios in \Tab{table:anisotropy_val}.

\bsp	
\label{lastpage}
\end{document}